\documentclass[11pt,a4paper]{article}

\usepackage{graphicx}
\usepackage{xcolor}
\usepackage{amsmath}
\usepackage{amssymb}
\usepackage{tikz}
\usepackage{ifthen}
\usepackage{natbib}

\usepackage{./myplots}
\newcommand{\degree}{\ensuremath{^\circ}}
\newcommand\thm{{\ensuremath{\theta_m}}}
\newcommand\pphi{{\ensuremath{\varphi}}}

\newcommand\dl{{\ensuremath{\delta_{l}}}}
\newcommand\dt{{\ensuremath{\delta_{t}}}}

\newcommand\dA{\mathrm{d}S}
\newcommand\dV{\mathrm{d}V}

\newcommand\ex{\vec{e}_x}
\newcommand\ez{\vec{e}_z}

\newcommand\vu{{\vec{u}}}
\newcommand\vw{{\vec{\omega}}}
\newcommand\pupt{{\frac{\partial\vu}{\partial\,t}}}

\newcommand\ct{{\ensuremath{c_t}}}
\newcommand\cta{{\ensuremath{c_t^m}}}
\newcommand\ctv{{\ensuremath{c_t^v}}}
\newcommand\cts{{\ensuremath{c_t^s}}}
\newcommand\cl{{\ensuremath{c_l}} }
\newcommand\cla{{\ensuremath{c_l^m}}}
\newcommand\clv{{\ensuremath{c_l^v}}}
\newcommand\cls{{\ensuremath{c_l^s}}}
\newcommand\mean[1]{\ensuremath{\overline{#1}}}
\newcommand\rms[1]{\ensuremath{#1'}}
\newcommand\ctm{{\ensuremath{\mean{c}_t}}} % mean value of thrust
\newcommand\clm{{\ensuremath{\mean{c}_l}}} % mean value of lift
\newcommand\ctr{{\ensuremath{\rms{c}_t}}} % rms value of thrust
\newcommand\clr{{\ensuremath{\rms{c}_l}}} % rms value of lift

\newcommand\im{\hphantom{-}}
\newcommand\om{\hphantom{1}}

\newcommand\phix{{\ensuremath{\phi_x}}}
\newcommand\phiz{{\ensuremath{\phi_z}}}

\let\ig\includegraphics
\tikzset{inner sep=0cm}
\title{On the aerodynamic forces \\ on heaving and pitching airfoils at low 
Reynolds number}

\author{M. Moriche, 
O. Flores, 
M. Garc\'{\i}a-Villalba \\
Departamento de Bioingenier\'{\i}a e Ingenier\'{\i}a Aeroespacial \\ 
Universidad Carlos III de Madrid \\
28911 Legan\'es, SPAIN} 
\date{}

\def\dummy{g}

\begin{document} %^ %^ %^ %^ %^ %^ %^ %^ %^ %^ %^ %^ %^ %^ %^ %^ %^ %^ %^ %^ %^

\maketitle

%
% JFM instruction: no more than 250 words
% currently: 233
\begin{abstract}
%-------------------------------------------------------------------------------
The influence that the kinematics of pitching and heaving 2D airfoils have on 
the aerodynamic forces is investigated using Direct Numerical Simulations and a 
force decomposition algorithm. 
Large amplitude motions are considered (of the order of one chord), with 
moderate Reynolds numbers and reduced frequencies of order $O(1)$, varying the 
mean pitch angle and the phase shift between the pitching and heaving motions. 
Our results show that the surface vorticity contribution  (viscous effects) to 
the aerodynamic force is negligible compared to the contributions from the body
motion (fluid inertia) and the vorticity within the flow (circulation). 
For the range of parameters considered here, the latter tends to be 
instantaneously oriented in the direction normal to the chord of the airfoil.
Based on the results discussed in the paper, a reduced order model for the 
instantaneous aerodynamic force is proposed, taking advantage of the force 
decomposition and the chord-normal orientation of the 
contribution from vorticity within the flow to the total aerodynamic
force.
The predictions of the proposed model are compared to those of a similar model 
from the literature, showing a noticeable improvement on the prediction of the 
mean thrust, and a smaller improvement on the prediction of mean lift and the 
instantaneous force coefficients.
\end{abstract}

%===============================================================================
%===============================================================================
%===introduction================================================================
%===============================================================================
%===============================================================================
\section{Introduction}
\label{sec:introduction}

%--------------------------------------------mavs-flapp-interes-----------------
Driven by the recent development of Micro Air Vehicles (MAVs), unsteady
aerodynamics of flapping wings has attracted the interest of the scientific
community during the past decades.
MAVs operating conditions are similar to those in which insects and small birds
fly: the Reynolds ($Re$) number of the flow is about $10$ to $10^4$ and the 
motion of the wings is characterized by moderate frequencies and high amplitudes
\citep{shyy:13}.
The maneuverability and performance of these animals is outstanding and, 
therefore, a deep insight in the aerodynamics of flapping flight is essential
to improve the design of MAVs. 
In particular, it is important to understand how aerodynamic forces are 
generated and to use this knowledge for the improvement of simplified force 
models.  
There is a broad literature on the aerodynamics of flapping wings as recently
reviewed by several authors 
\citep{rozhdestvensky:2003,platzer:2008,ellenrieder:2008,shyy:2010,shyy:13}. 
%_______________________________________________________________________________

%---------------------------------------pure-pitching---------------------------
In order to improve the understanding of flapping wing aerodynamics, scientists
have typically studied simplified configurations.
Numerous authors have studied the problem of a 2D airfoil in pure heaving 
motion, in which the airfoil oscillates vertically with a zero angle of attack
\citep{jones1997numerical,wang2000vortex,lewin2003modelling,
lua:2007,wei:2014,choi:2015}.
This simplified configuration is still a rich model where some of the main 
features of flapping flight are present, for example, the leading and trailing 
edge vortices.
The leading edge vortex (LEV) has been identified as the main lift enhancing 
mechanism of flapping wings \citep{ellington1996leading}.
%
% REBUTTAL-BEG
In fixed wing aerodynamics, the generation of an LEV produces a high lift
plateau for a short time span followed by a sudden drop of the aerodynamic
force \citep{carr:1988}.
% REBUTTAL-END
%
This process is known as dynamic stall.
Conversely, flapping wings take advantage of the high lift generated during
the formation of the LEV by consecutively generating an LEV in each stroke.
With this cyclic mechanism, the wing experiences the high transient lift from 
the generation of an LEV and avoids entering in the dynamic stall region.
%
%--------------------------------------Wang-2000-Vortex-------------------------
\citet{wang2000vortex} studied a heaving airfoil at $Re=1000$ by direct 
numerical simulations (DNS).
She found that tuning the motion parameters so that the time scales
of the motion and the LEV are similar, results in an optimal performance of the
airfoil in terms of the aerodynamic forces.
%
%--------------------------------------Lewing-2003-Modelling--------------------
% REBUTTAL-BEG
In an extensive numerical analysis on heaving airfoils at $Re=500$,
\citet{lewin2003modelling} explain how the interaction between the LEV and the 
trailing edge vortex (TEV) influences the propulsive efficiency of the airfoil.
% REBUTTAL-END
%------------------------------------------heaving-and-pitching-----------------
With the introduction of non-zero angle of attack, the airfoil may generate both
thrust and lift.
Heaving and pitching airfoils have also been studied extensively
\citep{anderson:1998,ramamurti:2001,read:2003,ashraf:2011,baik:2012,
widmann:2015}.
Pitching modifies the flow around the airfoil and can, for some combinations of 
the motion parameters, increase the net value of thrust and the propulsive 
efficiency.
%_______________________________________________________________________________

%-----------------------------------------Forces-decomposition------------------
In order to obtain a deeper insight in the generation of forces by flapping 
airfoils, several authors have proposed various methods to decompose the total 
aerodynamic force in different contributions 
\citep{chang1992potential,noca1999comparison,wu:2005}. 
These methods differ in the surface and volume integrals they involve, although 
it is possible to establish mathematical relations between them, see Appendix in
\citet{wang2014evaluation}. 
From a practical point of view it seems desirable to have a method where the 
terms are easy to compute and have a clear physical meaning.
%-------------------------------------------------------------------------------
Indeed, \citet{wang2014evaluation} proposes an approximate `simple lift formula'
decomposition, and compares it to the methods of \citet{noca1999comparison} and
\citet{wu:2005}, finding that  added mass and circulatory effects were the main
contributions to the aerodynamic force.
This result is in agreement with the recent work of \citet{martin2015vortex}, 
who employed the decomposition method proposed by \citet{chang1992potential} on
2D DNS data of a heaving airfoil at $Re=500$. 
\citet{martin2015vortex} also observed that the contribution to the aerodynamic
force of vortical structures which are a few chords away from the airfoil is 
negligible.
%_______________________________________________________________________________

%-------------------------------------------------------------------------------
The analysis of the aerodynamic forces in terms of their various contributions
might be used to predict the aerodynamic forces on a flapping wing from its 
geometry and kinematics.
New models to estimate the aerodynamic forces can be generated and, also, 
existing models can be improved.
Many such models exist since the pioneering work of \citet{wagner1925entstehung} 
and \citet{theodorsen1949general}, among others, and they have been recently 
reviewed by \citet{ansari:2006} and \citet{taha2012flight}.
For small amplitude motions at high $Re$ there is a complete theory based on
potential flow that predicts the aerodynamic forces produced on a thin airfoil 
\citep{wagner1925entstehung,theodorsen1949general}.
A less restrictive approach is provided by unsteady vortex lattice methods 
(UVLM). 
These methods present no restriction regarding the motion of the airfoil nor 
its geometry \citep{long2004object}.
However, since UVLM are also based on potential flow theory, they are not able
to capture leading edge separation. 
Therefore, these methods need modifications to include the contribution of LEVs,
as for example proposed by \citet{ansari:2006}.
Despite the fact that modified UVLM are computationally inexpensive compared to
other methods like DNS, their cost might still be too high to predict the 
forces on-the-fly in the small processors installed in MAVs.
This motivates the development of even simpler models, like the model proposed
by \citet{dickinson1999wing}.
This model is quasi-steady and uses algebraic expressions for the drag and lift
coefficients as a function of the instantaneous angle of attack, capturing the 
effect of the LEV in the force coefficients. 
The algebraic expressions in the model were calibrated using experimental data.
Another quasi-steady model was developed by \citet{pesavento2004falling} working
with free falling plates. They estimated the total aerodynamic force by 
separately modelling the added mass, circulatory and viscous effects.
In their work they propose an algebraic model for the circulation of an airfoil
in which high angles of attack and rotational effects are included.
A similar approach was followed by \citet{taha2014state} but including 
the effect of the wake using the Wagner function and neglecting viscous effects.
%_______________________________________________________________________________

%----------------------------------------and-this-is-how-we-do-=)---------------
With the aim of contributing to the improvement of simplified models for the 
aerodynamic forces, in this work we analyze DNS data of heaving and pitching 
airfoils, using the force 
decomposition method proposed by \citet{chang1992potential}.
Based on this analysis, we propose and test a simple model for the aerodynamic 
forces using the aforementioned force decomposition, and elements of the model
proposed by \citet{pesavento2004falling}. 
The database used in the present paper was first introduced in 
\citet{moriche:2015}, and some of the cases of this database were analyzed in 
\citet{moriche:2016}, with emphasis on the development of 3D instabilities. 
%_______________________________________________________________________________

%-----------------------------------------------------------Paper-Structure-----
The paper is organized as follows. In section \ref{sec:meth} the details of the 
numerical method, computational setup and force decomposition method are 
presented.
In section \ref{sec:forces} the complete database is described in terms of mean 
and rms total aerodynamic force coefficients.
In section \ref{sec:ref} the force decomposition method is applied to a 
reference case and the main contributions to the force are modelled.
After that, we extend the analysis to a subset of cases from the database in 
section \ref{sec:ext}, and to the whole database in section \ref{sec:tot}.
Finally, some conclusions are provided in section \ref{sec:concl}.
%_______________________________________________________________________________

%---------------------------------------------------------numerical-method------
\section{Numerical method}
\label{sec:meth}

%----------------------------------------what-we-do-with-TUCAN------------------
We present DNS of the flow around a symmetric airfoil NACA 0012 in heaving and 
pitching motion.
The Reynolds number of the flow based on the airfoil chord $c$ and the free 
stream velocity $U_\infty$ is $Re = c\,U_\infty/\nu = 1000$, where $\nu$ is the
kinematic viscosity of the fluid.
The simulations have been performed using TUCAN, a second order finite 
differences code that solves the Navier-Stokes equations for an incompressible 
flow \citep{moriche:2017}. 
The presence of the body is modelled by the direct forcing immersed boundary 
method proposed by \citet{uhlmann2005immersed}.
%_______________________________________________________________________________

\begin{figure}
\begin{center}
\begin{tikzpicture}
\coordinate(O) at (0., 0.);
\node(A) at (O){\ig[trim=0 0 0 0, clip, width=0.30\textwidth]{fig1a.eps}};
\node(B) at (A.east)[right,scale=0.8]{% Computational setup sketch
\def\myboxwest{
\begin{minipage}{2cm}
\centering
\begin{align}
 u &=  U_\infty \notag \\
 w &= 0          \notag
\end{align}
\end{minipage}
}
\def\myboxeast{
\begin{minipage}{2cm}
\begin{center}
\begin{align}\notag
\frac{\partial u}{\partial t}+ u_c \frac{\partial u}{\partial x}&=0\\
\notag
\frac{\partial w}{\partial t}+ u_c \frac{\partial w}{\partial x}&=0
\end{align}
\end{center}
\end{minipage}
}

\def\myboxfreeslip{
\begin{minipage}{2cm}
\centering
\begin{equation}\notag
\frac{\partial u}{\partial z} = 0 \,\text,\, w = 0  
\end{equation}
\end{minipage}
}

\begin{tikzpicture}
\def\chord{0.2}
\newlength\chordl
\setlength{\chordl}{0.25cm}
\coordinate(O) at  ( -7.5\chordl, 0.);
\coordinate(CO) at  ( 0., 0.);
\coordinate(LC) at (-12.5\chordl, -7.5\chordl);
\coordinate(RC) at ( 12.5\chordl,  7.5\chordl);
\fill[opacity=0.3]  ([xshift=0\chordl,yshift=-\chordl]O) --
         ([xshift=1\chordl,yshift=-\chordl]O) --
         ([xshift=1\chordl,yshift=+\chordl]O) --
         ([xshift=0\chordl,yshift=+\chordl]O);
\draw (LC) rectangle (RC);
\fill[color=blue] plot[xshift=-7.5\chordl,scale=\chord] file{./data/NACA0012.dat} -- cycle;
\draw[color=red] plot[xshift=-7.5\chordl,scale=\chord] file{./data/NACA0012.dat} -- cycle;
\node(west)  at (LC|-CO)[left ]{\myboxwest};
\node(east)  at ([xshift=0.2cm]RC|-CO)[right]{\myboxeast};
\node(north) at ([yshift=0.2cm]CO|-RC)[above]{\myboxfreeslip};
\node(south) at (CO|-LC)[below]{\myboxfreeslip};
\draw[-latex] (O)--([xshift=2\chordl]O);
\draw[-latex] (O)--([yshift=2\chordl]O);
\node(LX) at ([xshift=2\chordl]O)[below right]{$x$};
\node(LY) at ([yshift=2\chordl]O)[above left ]{$z$};

\draw[latex-latex] ([yshift=-1.00\chordl]LC|-O) -- ([yshift=-1.00\chordl]O);
\draw[latex-latex] ([yshift=-1.00\chordl,xshift=\chordl]O) -- ([yshift=-1.00\chordl]RC|-O);
\draw[latex-latex] ([xshift=5.5\chordl]LC) -- ([xshift=5.5\chordl,yshift=6.5\chordl]LC);
\node(AUX1) at ([yshift=-0.2\chordl,xshift=-3.0\chordl]O) {$5c$};
\node(AUX2) at ([yshift=-0.2\chordl,xshift=9.5\chordl]O) {$19c$};
\node(AUX3) at ([xshift=0.6\chordl,yshift=-4.25\chordl]O)[right]{$6.5c$};
\end{tikzpicture}};
\setcounter{myplotslabel}{1}
\mylabel{B}{-1.3in}{0.1in}
\mylabel{B}{0.6in}{0.1in}
\end{tikzpicture}
\caption{a) Sketch of the heaving and pitching motion of the airfoil. b) Sketch
of the computational domain. $u$ and $w$ are the velocities in the $x$ and $z$
directions, respectively and $u_c$ is the convective velocity. The airfoil is
represented in red and the region in which the motion takes place in light gray.
\label{outline}}
\end{center}
\end{figure}
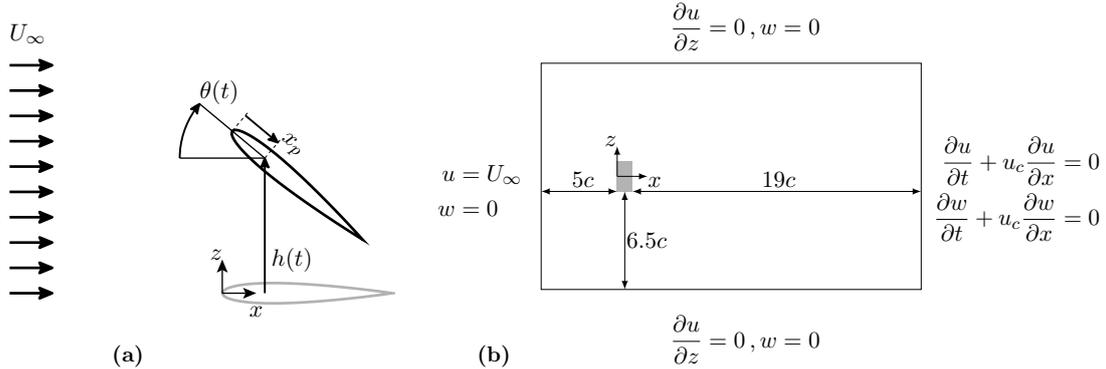

%----------------------------------------problem-definition---------------------
The prescribed heaving and pitching motion of the airfoil is given by
\begin{subequations}\label{eq:motion}
\begin{align}
     h(t) &=                 h_0 \cos(2\pi\,f\,t       )    \text{,} \\
\theta(t) &= \theta_m + \theta_0 \cos(2\pi\,f\,t + \varphi) \text{,}
\end{align}
\end{subequations}
where $h_0$ and $\theta_0$ are the heaving and pitching amplitude, respectively,
$\theta_m$ is the mean pitch value, $\varphi$ is the phase shift between the
heaving and pitching motion and $f$ is the frequency of the motion.
The pitching motion is a rotation around a point located at a distance $x_p$
from the LE, as it can be observed in the sketch shown in figure \ref{outline}a.
The set of non-dimensional parameters which define the problem ($h_0/c,\thm, 
\theta_0, \phi, 2\pi\,f\,c/U_\infty, x_p/c, \rho\,U_\infty/\nu$) results in a
large parametric space, so we only vary two of them.
The Reynolds number is $Re=1000$ and the pivoting point is located at the 
quarter of the chord ($x_p = c/4$).
The heaving amplitude is $h_0 = c$, the pitching amplitude is $\theta_0=30
\degree$ and the reduced frequency \footnote{Note that there is some ambiguity 
in the literature about the definition of the reduced frequency.
Some authors define it as $k=2\pi fc/U_\infty$, while others as 
$k=\pi fc/U_\infty$.
Here we have chosen the former.} is $k = 2\pi\,f\,c/U_\infty = 1.41$, resulting 
in a period of oscillation $T = 4.44\,c/U_\infty$.
The effect of the mean pitch angle and the phase shift between heaving and 
pitching are explored, varying \thm{} in the range $0\degree$ to $20\degree$ in 
steps $\Delta \thm{} = 10\degree$ and \pphi{} from $30\degree$ to $130\degree$ 
in steps $\Delta \pphi = 20\degree$, resulting in a database of $18$ 
simulations.
%_______________________________________________________________________________

%-------------------------------------------------------------------------------
All simulations are performed in a computational domain of dimensions $25c$ x 
$15c$ in the streamwise and vertical directions, respectively.
The resolution used in this study is $128$ points per chord, yielding a total of
$3200 \times 1920$ grid points in the streamwise and vertical directions,
respectively.
% 
% REBUTTAL BEG
This resolution has been selected based on a grid refinement study for a NACA 
0012 at $Re=1000$ set in heaving and pitching motion (see appendix 
\ref{sec:A1}).
% REBUTTAL END
The free stream condition is modelled by an inflow velocity $U_\infty$ at the 
inlet boundary, located $5c$ upstream of the airfoil's leading edge.
The outflow is modelled with an advective boundary condition at the outlet, 
located $19c$ downstream of the airfoil's trailing edge.
A free slip boundary condition is imposed at the lateral boundaries (see figure 
\ref{outline}b).
%_______________________________________________________________________________

%-------------------------------------------------------------------------------
The total aerodynamic force $\vec{F}$ is decomposed using the algorithm proposed
by \citet{chang1992potential} and recently used by \citet{martin2015vortex}.
The total aerodynamic force components in the streamwise ($x$) and vertical 
($z$) directions are expressed as
\begin{subequations} \label{eq:force_decomp}
\begin{equation} \label{eq:force_decomp_fx}
F_x = -\rho\int\limits_S \frac{\phix}{U_\infty} \pupt \cdot \vec{n}\, \dA 
      +\frac{\rho}{2} \int\limits_S |\vu|^2 \vec{n} \cdot \vec{e}_x\, \dA
      -\rho\int\limits_V \left( \vu \times \vw \right) \cdot
        \frac{\nabla\phix}{U_\infty} \, \dV 
      +\mu\int\limits_S  (\vw \times \vec{n}) \cdot \left(\frac{\nabla \phix}{U_\infty} + \vec{e}_x\right)
                                                               \,\dA \text{,}
\end{equation}
\begin{equation} \label{eq:force_decomp_fz}
F_z = -\rho\int\limits_S \frac{\phiz}{U_\infty} \pupt \cdot \vec{n} \,\dA 
      +\frac{\rho}{2}  \int\limits_S |\vu|^2 \vec{n}  \cdot \vec{e}_z\,\dA
      -\rho\int\limits_V \left( \vu \times \vw \right) \frac{\cdot\nabla\phiz}{U_\infty}  \,\dV 
      +\mu\int\limits_S (\vw \times \vec{n}) \cdot \left(\frac{\nabla \phiz}{U_\infty} + \vec{e}_z\right) 
                                                               \,\dA \text{,}
\end{equation}
\end{subequations}
where $\vu$ is the velocity of the flow, $\vw$ is the vorticity, $\mu$ is the
dynamic viscosity of the fluid, $S$ the surface of the airfoil, $V$ the fluid 
domain, $\vec{n}$ the unitary vector normal to the surface of the airfoil, 
pointing towards the fluid, and $\vec{e}_x$ and $\vec{e}_z$ are the unitary 
vectors in the $x$ and $z$ directions, respectively.
%_______________________________________________________________________________
%
The auxiliary potentials \phix{} and \phiz{} that appear in equation 
\eqref{eq:force_decomp} depend only on the geometry of the airfoil and on the 
directions in which they are computed.
For details of the calculations of these potentials, the reader is referred to 
the Appendix \ref{sec:rotation}.
%_______________________________________________________________________________

%----------------------------------------------------group-terms----------------
Following \cite{chang1992potential}, we group the terms of equation
\eqref{eq:force_decomp} to identify three different contributions to the 
aerodynamic forces, 
\begin{equation} 
\vec{F} = \vec{F}^m + \vec{F}^v + \vec{F}^s.
\label{eq:vecforce_decomp}
\end{equation}
The first two terms of the right hand side in equation \eqref{eq:force_decomp} 
are the contribution due to the motion of the body, $\vec{F}^m$. 
The contribution of the vorticity within the flow, $\vec{F}^v$,  is given by the
third term.
Finally, the surface vorticity contribution, $\vec{F}^s$, is the last term of
equation \eqref{eq:force_decomp}.
%_______________________________________________________________________________

%---------------------------------------------------advantages------------------
The decomposition described in equation \eqref{eq:force_decomp} presents some 
advantages with respect to other algorithms found in the literature.
First, the contribution of the body motion is calculated with surface integrals 
which only involve the velocity of the flow and the auxiliary potential 
functions, both on the surface of the airfoil.
Hence, the contribution for $\vec{F}^m$ is prescribed by the geometry and the
kinematics of the airfoil, and can be computed a priori (see section 
\ref{sec:ref}).
Second, the only time derivative of the fluid velocity in equation 
\eqref{eq:force_decomp} appears in a surface integral, so that $\pupt$ can be 
evaluated from the kinematics of the airfoil.
This means that this force decomposition algorithm can be applied to isolated 
snapshots of the velocity field, e.g. obtained from particle image velocimetry 
measurements.
Finally, the integrand of the contribution of the vorticity within the flow can 
be interpreted as a force density, allowing a direct evaluation of how specific 
vortices within the flow contribute to the total aerodynamic force.
%_______________________________________________________________________________

%-----------------------------------------------ct-cl-definitions---------------
The aerodynamic forces in equation \eqref{eq:force_decomp} can be made 
dimensionless using the density $\rho$, the free stream velocity $U_\infty$ and
the chord $c$, resulting in the non-dimensional coefficients of thrust ($\ct$)  
and lift ($\cl$)
\begin{equation}\label{eq:ctcl}
\ct = \frac{  -2\,F_x}{\rho\,U_\infty^2\,c} \text{,} \quad
\cl = \frac{\im2\,F_z}{\rho\,U_\infty^2\,c} \text{.}
\end{equation}
Analogous to equation \eqref{eq:vecforce_decomp}, the total non-dimensional
force coefficients are split into 
\begin{subequations}\label{eq:force_decomp_cf}
\begin{align}
\ct &= \cta + \ctv + \cts,   \\
\cl &= \cla + \clv + \cls. 
\end{align}
\end{subequations}
%-----------------------------------------------thrust-lift-density-------------
Furthermore, we define the spatial density of thrust \dt{} and lift \dl{} as the
integrand of the contribution of the vorticity within the flow to the total 
aerodynamic force in equation \eqref{eq:force_decomp}.
After non-dimensionalization, the definitions of \dt{} and \dl{} read
\begin{subequations}\label{eq:vf}
\begin{align}
\dt &= \frac{\im2\left(\vu \times \vw \right)\cdot\nabla\phix}{ U_\infty^3/c}
\text{,}\label{eq:vft} \\
\dl &= \frac{-2\left(\vu \times \vw \right)\cdot\nabla\phiz}{ U_\infty^3/c}
\text{.}\label{eq:vfl}
\end{align}
\end{subequations}
%-----------------------------------------------mean-rms-values-----------------
%
Also, to support the following discussions we define the average  of any time 
dependent variable $\dummy(t)$, as
\begin{equation}
\mean{\dummy} = \frac{1}{T_{av}}\int_0^{T_{av}} \dummy(t) \mathrm{d}t \text{,}
\end{equation}
where $T_{av}$ is the time span for the averaging process.
Finally, the root mean square (rms)  of $\dummy(t)$ is defined as 
\begin{equation}
\rms{\dummy} =\sqrt{\frac{1}{T_{av}}\int_0^{T_{av}}
\left(\dummy(t)^2-\mean{\dummy}^2\right) \mathrm{d}t}\text{.}
\end{equation}

%===============================================================================
%====Results====================================================================
%===============================================================================
\section{Aerodynamic forces}
\label{sec:forces}

\begin{table}
\begin{center}
\begin{tabular*}{\textwidth}{@{\extracolsep{\fill}}cccccccccc} 
Case & \thm($\degree$) & \pphi($\degree$)& $T_{av}/T$ & Periodicity &  \mean{\ct} &      \rms{\ct} &     \mean{\cl} &      \rms{\cl} &         $\eta$  \\
\hline                                               
A030 & $\om        0$ & $\om       30$ & 10 & A & $     -0.4382$ & $      0.6144$ & $      0.0041$ & $      3.3181$ & $0$             \\ 
A050 & $\om        0$ & $\om       50$ &  1 & P & $\im   0.1566$ & $      0.7184$ & $      0.0000$ & $      2.6276$ & $      0.0834$  \\ 
A070 & $\om        0$ & $\om       70$ &  1 & P & $\im   0.8062$ & $      0.7550$ & $      0.0000$ & $      2.4396$ & $      0.3374$  \\ 
A090 & $\om        0$ & $\om       90$ &  1 & P & $\im   0.9957$ & $      0.9751$ & $      0.0000$ & $      2.7164$ & $      0.3644$  \\ 
A110 & $\om        0$ & $         110$ &  1 & P & $\im   1.0439$ & $      1.2878$ & $      0.0000$ & $      3.7486$ & $      0.2933$  \\ 
A130 & $\om        0$ & $         130$ &  1 & P & $\im   0.9582$ & $      1.4988$ & $      0.0046$ & $      5.2013$ & $      0.1937$  \\ 
B030 & $          10$ & $\om       30$ &  5 & A & $     -0.3202$ & $      0.7715$ & $      0.2662$ & $      3.0791$ & $0$\\              
B050 & $          10$ & $\om       50$ &  1 & P & $\im   0.0972$ & $      0.7235$ & $      0.4163$ & $      2.6734$ & $      0.0519$  \\ 
B070 & $          10$ & $\om       70$ &  1 & P & $\im   0.5787$ & $      0.7017$ & $      0.8312$ & $      2.7040$ & $      0.2486$  \\ 
B090 & $          10$ & $\om       90$ &  1 & P & $\im   0.7245$ & $      0.9224$ & $      1.5507$ & $      2.7743$ & $      0.2620$  \\ 
B110 & $          10$ & $         110$ &  1 & P & $\im   0.8635$ & $      1.2008$ & $      1.6032$ & $      3.7622$ & $      0.2348$  \\ 
B130 & $          10$ & $         130$ &  1 & P & $\im   0.7547$ & $      1.4351$ & $      1.6678$ & $      5.0979$ & $      0.1539$  \\ 
C030 & $          20$ & $\om       30$ & 10 & A & $     -0.9957$ & $      1.1545$ & $      1.2519$ & $      3.1465$ & $0$\\              
C050 & $          20$ & $\om       50$ &  1 & P & $     -0.8843$ & $      0.7999$ & $      2.2511$ & $      2.3009$ & $0$\\              
C070 & $          20$ & $\om       70$ & 15 & A & $     -0.0468$ & $      0.5790$ & $      1.7455$ & $      2.7850$ & $0$\\              
C090 & $          20$ & $\om       90$ &  2 & D & $     -0.1419$ & $      0.9511$ & $      2.7850$ & $      2.9327$ & $0$\\              
C110 & $          20$ & $         110$ &  1 & P & $\im   0.1425$ & $      1.0816$ & $      3.0268$ & $      3.7850$ & $      0.0392$  \\ 
C130 & $          20$ & $         130$ &  2 & D & $\im   0.1132$ & $      1.2640$ & $      3.1992$ & $      4.8421$ & $      0.0242$  \\ 
\end{tabular*}
\end{center}
\caption{Motion parameters and integrated values of non-dimensional
force coefficients of thrust and lift of all the cases.
The periodicity of the flow is indicated with P for periodic,
D for periodic with period $2T$ and A for aperiodic.
\label{tab:clct_all}}
\end{table}

\begin{figure}
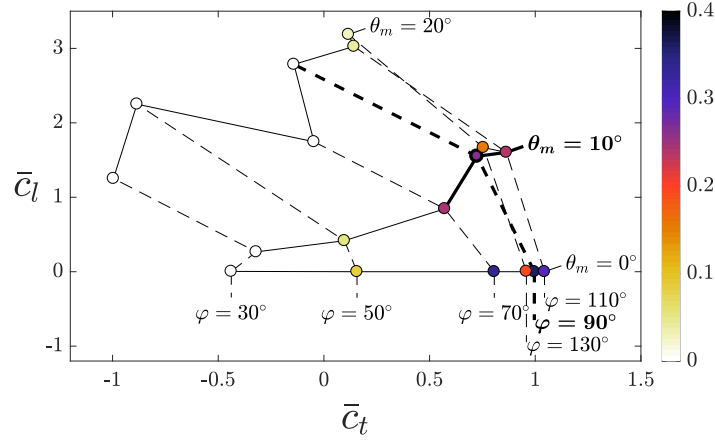

\begin{center}
\ig[scale=0.5]{fig2.eps}
\caption{$\clm$ vs. $\ctm$ for all cases in the database.
The color corresponds to the propulsive efficiency, $\eta$.
Solid (dashed) lines connect cases with constant mean pitch value $\theta_m$
(phase shift $\varphi$).
\label{clct_all}}
\end{center}
\end{figure}

%-------------------------------------------------------------------------------
Table \ref{tab:clct_all} shows the averaged and rms values of thrust and lift 
coefficients for the 18 cases described in section \ref{sec:meth}, together with
the propulsive efficiency
\begin{equation}
\eta=\frac{T_{av}\,\mean{F}_x\,U_\infty}{
\int_{0}^{T_{av}}\left(F_z\,\dot{h}+M_{y,c/4}\,\dot{\theta}\right)
\mathrm{d}t} \text{,}
\end{equation}
where $M_{y,c/4}$ is the pitching moment about the quarter of the chord
(coincident with the hinge point).
Note that the propulsive efficiency is set to zero for cases with net drag, 
avoiding meaningless negative values.
The cases are identified by a letter (A, B or C) related to the value of \thm{} 
($0\degree$, $10\degree$ or $20\degree$, respectively), followed by three digits
that correspond to the value of $\pphi$.
Table \ref{tab:clct_all} also shows that most of the cases present the same 
periodicity in the flow and forces as in the motion, with period 
$T U_\infty/c=4.44$.
However, there are two cases with a doubling period phenomena and four cases 
that are aperiodic.
Therefore, the time span for the averaging $T_{av}$ for the mean and rms 
coefficients of each case has been selected accordingly, as seen in the 
fourth column in table \ref{tab:clct_all}. 
The appearance of aperiodic behavior has been previously observed by other 
authors, for example by \citet{lewin2003modelling} in pure heaving cases.
%-------------------------------------------------------------------------------

%-------------------------------------------------------------------------------
A graphical representation of  the data in table \ref{tab:clct_all} is provided
in figure \ref{clct_all}, where each case is represented by a point in the 
\ctm{}, \clm{} phase space.
It can be seen that when the phase shift is fixed, an increase of the mean pitch
value results in an increase of lift and a reduction of thrust (dashed lines in 
figure \ref{clct_all}).
When the mean pitch value is fixed (solid lines in figure \ref{clct_all}), 
\ctm{} and \clm{} tend to increase with $\varphi$. 
For moderate mean pitch angles, maximum propulsive efficiency is achieved for 
$\varphi\approx 90\degree$, while maximum force coefficients are obtained for 
slightly larger phase shifts, $\varphi \approx 110\degree$, consistent with the
results of \citet{anderson:1998}.
More specifically, for a mean pitch value \thm{} equal to $0\degree$ and 
$10\degree$, the maximum thrust is obtained for a phase shift 
$\pphi=110\degree$.
The highest propulsive efficiency is $36\%$, obtained for $\thm=0\degree$ and a
phase shift $\pphi=90\degree$.
Also, cases with $\thm=10\degree$ and $\pphi=90-110\degree$ yield relatively
high $\mean{\ct}$ and $\mean{\cl}$, with propulsive efficiencies higher than 
$20\%$.
Finally, note that for $\thm=20\degree$ the propulsive efficiencies are 
considerably lower, with higher lift coefficients and a tendency to lose 
periodicity.
%_______________________________________________________________________________

%-------------------------------------------------------------------------------
We select case B090 as a reference case to perform a more detailed analysis. 
This choice is motivated by the fact that B090 provides both thrust and lift 
with a relatively high propulsive efficiency, $\eta= 26\%$, so it is interesting
in terms of aerodynamic performance.
Also, by varying the phase shift or the mean pitch value of case B090, a subset
of cases from the database can defined, which allows the analysis of the 
influence of the motion parameters on the aerodynamic forces.
This subset of cases is represented in figure \ref{clct_all} with thicker solid 
and dashed lines, and it includes cases A090, C090, B070 and B110.
%_______________________________________________________________________________

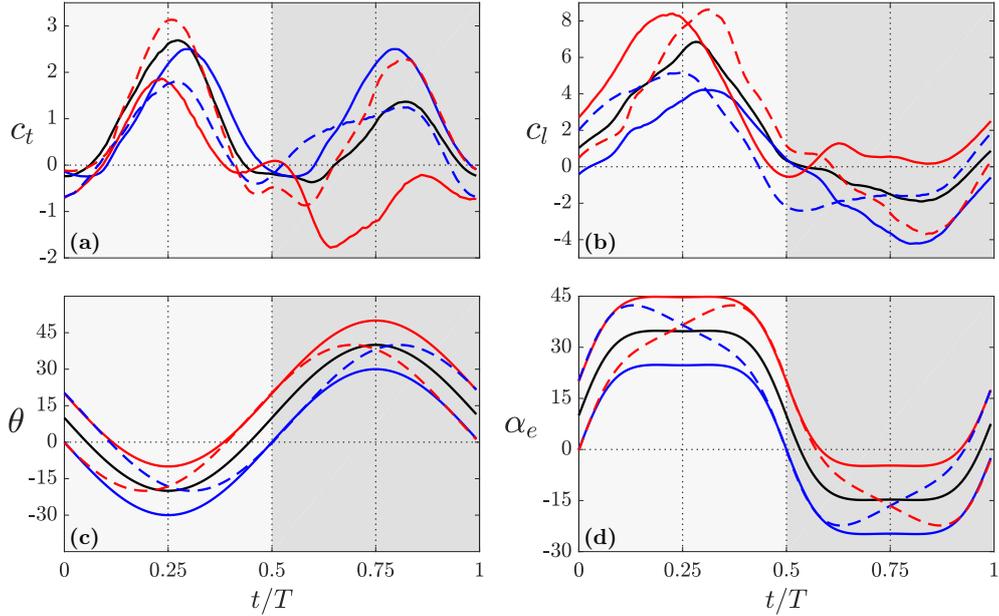
\begin{figure}
\begin{center}
\begin{tikzpicture}
\coordinate(O) at (0cm, 0cm);
\node[](T1) at (O)              {\ig[scale=0.5]{fig3a.eps}};
\node[](L1) at (T1.east) [right]{\ig[scale=0.5]{fig3b.eps}};
\node[](AE1)at (T1.south)[below]{\ig[scale=0.5]{fig3c.eps}};
\node[](AE2)at (L1.south)[below]{\ig[scale=0.5]{fig3d.eps}};
\setcounter{myplotslabel}{1}
\myoplabelar{T1}{2.2572in}{1.4332in}
\myoplabelar{L1}{2.2572in}{1.4332in}
\myoplabelar{AE1}{2.2572in}{1.4332in}
\myoplabelar{AE2}{2.2572in}{1.4332in}
\end{tikzpicture}
\caption{%\protect\input{figtab/\tbma/\tbma_1000_data/\tbma_1000.glb}
a) Thrust and b) lift coefficient of the selected cases with respect to the 
reference case B090 (\lineSymbolRGB[solid]{0}{0}{0}{0.012in}{none}{0in}). 
c) Pitch angle of the airfoil and d) effective angle of attack (in degrees). 
Cases where \thm{} is modified are represented with solid lines, 
case A090 (\lineSymbolRGB[solid]{0}{0}{1}{0.012in}{none}{0in}) and 
case C090 (\lineSymbolRGB[solid]{1}{0}{0}{0.012in}{none}{0in}), and cases where 
\pphi{} is modified are represented with dashed lines, 
case B070 (\lineSymbolRGB[dashed]{0}{0}{1}{0.012in}{none}{0in}) and 
case B110 (\lineSymbolRGB[dashed]{1}{0}{0}{0.012in}{none}{0in}).
The downstroke (upstroke) is indicated by a light (dark) grey background.
\label{forcesdecomp_focus_TOTAL}}
\end{center}
\end{figure}

\def\mf{\ref{forcesdecomp_focus_TOTAL}}
%-%%/////////////////////////////////////////////////////////////// TOTAL COMPONENT
In the reference case B090, the net thrust and lift ($\ctm=0.72$, $\clm=1.55$)
are of the same order of magnitude as the rms of the force fluctuations
($\ctr=0.92$, $\clr=2.77$).
This reflects that the oscillatory component of the force is as important as its
mean value.
For case B090, figures \mf{}a and b show the time evolution of $\ct$ and $\cl$, 
respectively, during one period.
Two peaks of thrust (figure \mf{}a) are generated during the period, one in the 
downstroke and one in the upstroke, with a larger magnitude in the former.
Regarding the lift (figure \mf{}b), the large lift generated in the downstroke 
is  partially counteracted by the negative lift produced in the upstroke. 
%_______________________________________________________________________________

\def\mf{\ref{forcesdecomp_focus_TOTAL}} % FIGURE MOSTLY COMMENTED 
%-------------------------------------------------------------------------------
The influence of the mean pitch angle is analyzed in detail comparing cases 
A090, B090 and C090, which have a constant phase shift $\pphi$ of $90\degree$ 
and a mean pitch value \thm{} of $0\degree$, $10\degree$ and $20\degree$, 
respectively.
If the mean pitch value \thm{} is set to zero (case A090), the performance of 
the airfoil is improved with respect to case B090 ($\thm=10\degree$) in terms of
net thrust, but at the cost of producing zero net lift.
The $\ctm$ generated by case A090 is increased by $37.5\%$ with respect to B090
and the propulsive efficiency by $40\%$.
Conversely, if the mean pitch value \thm{} is increased to $20\degree$ (case 
C090), a larger lift is generated ($\clm = 2.8$), at the cost of producing a 
small net drag ($\ctm = -0.15$).
Concerning the variation of the force during the period, rms values of lift and
thrust are comparable to mean values, as in the reference case. Moreover, $\ctr$
and  $\clr$ are roughly insensitive to variations of the mean pitch value, with 
$\ctr\approx0.9$ and $\clr\approx2.8$ for cases A090, B090 and C090.
In terms of the instantaneous forces, figure \mf{}a shows that, during the 
downstroke, the thrust generated in case A090 ($\thm=0\degree$) is similar to 
the thrust generated in case B090 ($\thm=10\degree$).
%-------------------------------------------------------------------------------
However, during the upstroke the thrust generated is notably larger in case A090
compared to case B090.
Conversely, case C090 ($\thm=20\degree$) generates less thrust during the whole 
period, with negative values of $\ct(t)$ during the upstroke, resulting in net 
drag.
The behavior of the instantaneous lift coefficient (figure \mf{}b) is roughly
the opposite, increasing with $\theta_m$ during the whole period, except in the
transition from downstroke to upstroke.
This transition is marked by a drop in $\cl$ associated to the detachment of the
LEV, which occurs earlier for the cases with higher $\thm$.
Overall, the variation of $\cl$ and $\ct$ with $\thm$ is consistent with an 
increase of the total force as $\thm$ increases, coupled with a change of the
orientation of that force, which is tilted backwards as \thm{} increases. 
%_______________________________________________________________________________

%-------------------------------------------------------------------------------
The effect of the phase shift on the aerodynamic forces is less intuitive. 
This effect is analyzed comparing cases B070, B090 and B110, which have a 
constant mean pitch value $\thm$ of $10\degree$ and a phase shift $\pphi$ of 
$70\degree$, $90\degree$ and $110\degree$, respectively.
A lag in the pitching motion ($\pphi<90\degree$, as in case B070) results in 
lower mean forces, specially for the mean lift coefficient. 
On the other hand, an advance of the pitching motion ($\pphi>90\degree$, as in
case B110) results in higher net thrust and lift. 
Besides the change on the mean values, the main effect of the phase shift on the
instantaneous force coefficients is an increase of the amplitude of the force
perturbations with $\varphi$. 
This can be observed both in figures \mf{}a and b, as well as in the $\ctr$ and
$\clr$ reported in table \ref{tab:clct_all}.  
Finally, it is clear from figure \mf{}a that $\varphi$ also modifies the time at
which the lift coefficients are maximum. This effect on the lift can be partly 
explained by the evolution of the effective angle of attack $\alpha_e = \theta 
- \mathrm{atan}(\dot{h}/U_\infty)$, shown in figure \mf{}d. 
For $\pphi>90\degree$ (B110), the peaks of $\alpha_e$ are delayed in time with
respect to B090.  As a consequence, the peaks of $c_l$ occur later for B110 than
for B090.
On the other hand, for $\pphi<90\degree$ (B070), the peaks of $\alpha_e$ and
$c_l$ are advanced with respect to B090. 
Note that a similar trend is not apparent in the times where $c_t$ is maximum,
which are fairly independent of the phase shift.
The reason for this is the importance of the pitch angle in the orientation of
the resulting aerodynamic force: if we assume that the aerodynamic forces are
mostly perpendicular to the airfoil (an assumption that will be justified 
later), then the variation in the pitch angle during the upstroke or downstroke
between cases B070, B090 and B110 results in a small effect for the lift (which
is multiplied by  $\cos \theta$),  but a larger effect on the thrust (which is
multiplied by $\sin \theta$).
Note that for these cases, the pitch angle, $\theta$, shown in figure 
\ref{forcesdecomp_focus_TOTAL}c, is such that $\cos \theta \sim 1$ and 
$\sin \theta \sim \theta$.
As a consequence, the lift is dominated by $\alpha_e$ while the thrust depends
on both, $\alpha_e$ and $\theta$. 
%_______________________________________________________________________________

%===============================================================================
% FORCE DECOMPOSITION 
\section{Force decomposition and modelling of case B090}
\label{sec:ref}

%--------------------------------------------------B090-force decomp------------
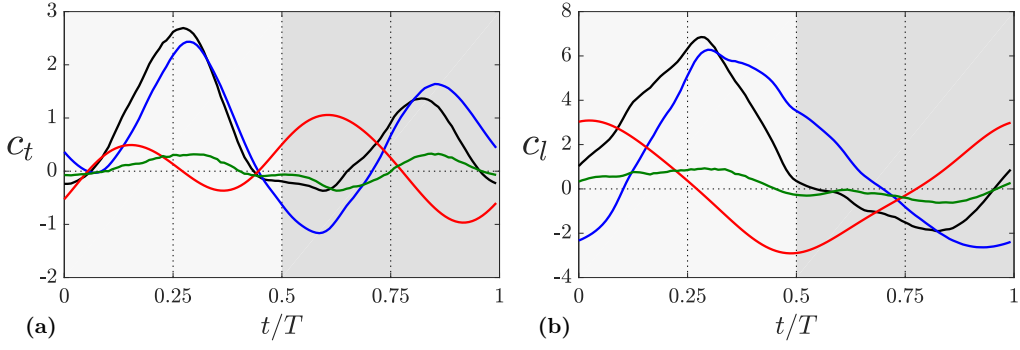
\begin{figure}
\begin{center}
\begin{tikzpicture}
\coordinate(O) at (0cm, 0cm);
\node[](A) at (O)[above left ]{\ig[scale=0.5]{fig4a.eps}};
\node[](B) at (O)[above right]{\ig[scale=0.5]{fig4b.eps}};
\setcounter{myplotslabel}{1}
\myoplabel{A}{2.4in}{1.7in}
\myoplabel{B}{2.4in}{1.7in}
\end{tikzpicture}
\caption{%\protect\input{figtab/\tbm/\tbm_1000_data/\tbm_1000.glb}
a) Thrust and b) lift coefficient of case B090. Curves represented correspond to
the total aerodynamic force ($c_l, c_t$: \lineSymbolRGB[solid]{0}{0}{0}{0.012in}{none}{0in})
and contributions from body motion 
($c_l^m, c_t^m$: \lineSymbolRGB[solid]{1}{0}{0}{0.012in}{none}{0in}), vorticity within the flow 
($c_l^v, c_t^v$: \lineSymbolRGB[solid]{0}{0}{1}{0.012in}{none}{0in}) and surface vorticity      
($c_l^s, c_t^s$: \lineSymbolRGB[solid]{0}{0.5}{0}{0.012in}{none}{0in}).
The downstroke (upstroke) is indicated by a light (dark) grey background.
\label{forcesdecomp_detail}}
\end{center}
\end{figure}
\def\mf{\ref{forcesdecomp_detail}}
%-----------------------------------------B090-decomposition-intro--------------
Following the procedure described in section \ref{sec:meth}, we decompose the
total aerodynamic force of case B090.
Figure \mf{} shows the evolution of the total thrust and lift, together with the
contributions from body motion, $\vec{F}^m$, vorticity within the flow,
$\vec{F}^v$, and surface vorticity, $\vec{F}^s$, during one period of case B090.
The main contribution to the total aerodynamic force corresponds to the
vorticity within the flow, with peak values of the same order of magnitude as 
the total value of the force.
Body motion has also an important role in the generation of force, producing 
peak values around half of the peak values of the total aerodynamic force.
Finally, surface vorticity (viscous effects) is the least important 
contribution, with peak values approximately ten times smaller than the total 
force peak values.
Therefore, in the following we will focus in the analysis of the contributions 
from body motion and the vorticity within the flow.
%_______________________________________________________________________________

%------------------------------------------Added-mass-analysis-B090-------------
We start with the contribution of the body motion to the total force, 
$\vec{F}^m$, which is the force produced by the fluid to counteract the motion 
of the airfoil.
This is easily observed in figure \mf{}b: when $\ddot{h}<0$, $\cla$ 
is a positive vertical force, and viceversa.
The thrust (figure \mf{}a) is influenced by both the vertical acceleration and
the projected area of the airfoil perpendicular to the streamwise direction.
For the motion parameters of this case ($\thm=10\degree$, $\pphi=90\degree$),
the projected area during the downstroke is smaller than the projected area 
during the upstroke, resulting in higher peaks of $\cta$ in the latter.
%_______________________________________________________________________________

\begin{figure}
\begin{center}
\begin{tikzpicture}
\coordinate(O) at (0.,0.);
\node(A) at (O)            {\ig[scale=0.5]{fig5a.eps}};
\node(B) at (A.east)[right]{\ig[scale=0.5]{fig5b.eps}};
\setcounter{myplotslabel}{1}
\mylabel{A}{0.3in}{0.3in}
\mylabel{B}{0.3in}{0.3in}
\end{tikzpicture}
\caption{Mean a) thrust and b) lift coefficients from body motion contribution 
for a flat plate with respect to the phase shift \pphi{}. 
The values shown correspond to different values of mean pitch angle,
$\thm = 0\degree$ (\lineSymbolRGB[solid]{0}{0.0}{1}{0.012in}{none}{0in}), 
$\thm =10\degree$ (\lineSymbolRGB[solid]{1}{0.0}{0}{0.012in}{none}{0in}) and 
$\thm =20\degree$ (\lineSymbolRGB[solid]{0}{0.6}{0}{0.012in}{none}{0in}).
\label{fig:bodymotion}}
\end{center}
\end{figure}
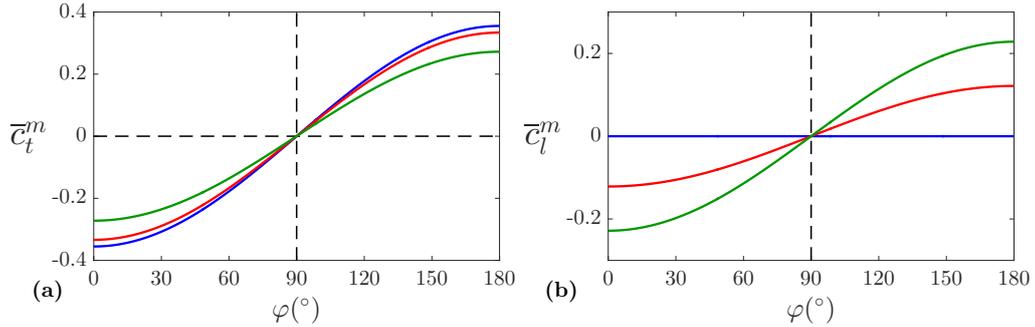

%----------------------------------------------------potential-theroy-added-mass
From a physical point of view, $\vec{F}^m$ is similar to the added-mass term of
the aerodynamic forces in unsteady potential flow. 
However, they are not exactly the same. 
This is shown here for a flat plate with $x_p = c/2$. According to 
\citet{sedov1965two}, the added mass forces for this configuration are 
\begin{subequations}\label{eq:added}
\begin{align}
c_t^a &= 
  \frac{\pi}{4} \frac{\ddot{h}}{U_\infty^2/c} \sin\left(2\theta\right)
+\frac{\pi}{2}\frac{\dot{h}}{U_\infty}
              \frac{\dot{\theta}}{U_\infty/c}\cos^2\left(\theta\right)
-\frac{\pi}{4} \left( 
  \frac{\dot{\theta}^2}{U_\infty^2/c^2} \cos\theta
 +\frac{\ddot{\theta}}{U_\infty^2/c^2}  \sin\theta
  \right),\\
c_l^a &=
  -\frac{\pi}{2} \frac{\ddot{h}}{U_\infty^2/c} \cos^2\theta
  + \frac{\pi}{4}\frac{\dot{h}}{U_\infty}
              \frac{\dot{\theta}}{U_\infty/c}\sin\left(2\theta\right)
+\frac{\pi}{4} \left( 
  \frac{\ddot{\theta}}{U_\infty^2/c^2}  \cos\theta
 -\frac{\dot{\theta}^2}{U_\infty^2/c^2} \sin\theta
  \right).
\end{align}
\end{subequations}
However, the corresponding expressions for the $c_t^m$ and $c_l^m$ are 
\begin{subequations}\label{eq:bm}
\begin{align}
\cta=\frac{\pi}{4}\frac{\ddot{h}}{U_\infty^2/c}\sin\left(2\theta\right),\\
\cla = -\frac{\pi}{2} \frac{\ddot{h}}{U_\infty^2/c} \cos^2\theta, 
\end{align}
\end{subequations}
where the analytical expressions for the auxiliary potentials of a flat plate 
are obtained from \citep{martin2015vortex}.
%-------------------------------------------------------------------------------
Note that while the first term in equation (\ref{eq:added}) is the same
appearing in equation (\ref{eq:bm}), the former has some extra terms depending
explicitly on $\dot{\theta}$ and $\ddot{\theta}$.
Moreover, for periodic motions the mean value of the added mass coefficients are 
$\overline{c}_l^a=\overline{c}_t^a=0$, while the mean value of the body motion
contributions are only zero if $\varphi=90\degree$. 
This can be observed in figure \ref{fig:bodymotion}, which shows $\ctm^m$ and 
$\clm^m$ as a function of $\varphi$ for $\theta_m = 0^\circ, 10^\circ$ and 
$20^\circ$. 
Interestingly, the behaviour of $\ctm^m$ and $\clm^m$ in figure 
\ref{fig:bodymotion} is consistent with the trends shown in figure 
\ref{clct_all}, except for maybe the behavior for $\varphi\gtrsim 110 \degree$.
%_______________________________________________________________________________

%-------------------------------------------------------------------------------
Finally, from the point of view of modelling, it is interesting to note that 
$\vec{F}^m$ can be computed a priori, since the velocity on the surface is known
from the airfoil's kinematics.
Note also that, as explained in appendix \ref{sec:rotation}, the auxiliary
potentials can be computed on a reference frame fixed to the airfoil, and then
rotated and translated appropriately to account for the motion of the airfoil. 
For reference, the auxiliary potentials for the NACA 0012 airfoil used in this
work are provided in figure \ref{fig:phiatfoil} in appendix \ref{sec:rotation}.
%_______________________________________________________________________________

%_______________________________________________________________________________
%-------------------------------------------------------------------------------

%-------------------------------------------volumetric-force--------------------
Now, we turn our attention to the contribution that the vorticity within the 
flow has on the total aerodynamic force of case B090 (blue lines in figure 
\ref{forcesdecomp_detail}).
It is clear that the (positive) peaks of total thrust and lift are dominated by 
the contribution of the vorticity within the flow.
Also, the contribution of the vorticity within the flow is maximum when the 
vertical velocity of the airfoil $\dot{h}$ and the effective angle of attack 
$\alpha_e$ are maximum.
Therefore, two peaks of positive thrust are observed in figure \mf{}a, slightly
lagged with respect to the middle of the downstroke ($t/T=0.25$) and  the middle
of the upstroke ($t/T=0.75$), respectively.
Regarding the lift, the peak of the instantaneous force coefficients are also
slightly lagged. 
The peak of force generated around the middle of the downstroke ($t/T=0.25$) is
positive and the one generated around the middle of the upstroke ($t/T=0.75$)
is negative.
The asymmetry introduced by the mean pitch angle $\thm =10\degree$ results in
lower peak values in the upstroke compared to the downstroke, which is 
detrimental for the thrust, but favourable for the lift.
%_______________________________________________________________________________

%/////////////////////////////////////////////////////////////flow visualization
\begin{figure}
\begin{center}
\begin{tikzpicture}
% D: downstroke
% W,T,L: Vorticity, Thrust, Lift
% N: NT/8
\coordinate(O) at (0cm, 0cm);
\node(W0) at (O)              {\ig[scale=0.5]{fig6a.eps}};
\node(T0) at (W0.east) [right]{\ig[scale=0.5]{fig6b.eps}};
\node(L0) at (T0.east) [right]{\ig[scale=0.5]{fig6c.eps}};

\node(W2) at ([yshift=0.2cm]W0.south)[below]{\ig[scale=0.5]{fig6d.eps}};
\node(T2) at (W2.east)               [right]{\ig[scale=0.5]{fig6e.eps}};
\node(L2) at (T2.east)               [right]{\ig[scale=0.5]{fig6f.eps}};
                                                                 
\node(W4) at ([yshift=0.2cm]W2.south)[below]{\ig[scale=0.5]{fig6g.eps}};
\node(T4) at (W4.east)               [right]{\ig[scale=0.5]{fig6h.eps}};
\node(L4) at (T4.east)               [right]{\ig[scale=0.5]{fig6i.eps}};
                                                                 
\node(W6) at ([yshift=0.2cm]W4.south)[below]{\ig[scale=0.5]{fig6j.eps}};
\node(T6) at (W6.east)               [right]{\ig[scale=0.5]{fig6k.eps}};
\node(L6) at (T6.east)               [right]{\ig[scale=0.5]{fig6l.eps}};

% cycle symbol                                                     
\drawcyclep{[xshift=0.5cm]L0.east}{0}
\drawcyclep{[xshift=0.5cm]L2.east}{2}
\drawcyclep{[xshift=0.5cm]L4.east}{4}
\drawcyclep{[xshift=0.5cm]L6.east}{6}

% colorbars
\node(CBW) at ([yshift=-0.00cm]W0.north)[above]{\ig[scale=0.5]{fig6bar1.eps}};
\node(CBL) at ([yshift=-0.00cm]L0.north)[above]{\ig[scale=0.5]{fig6bar2.eps}};
\node(CBT) at ([yshift=-0.00cm]T0.north)[above]{\ig[scale=0.5]{fig6bar3.eps}};

% labels of vorticity
\setcounter{myplotslabel}{1}
\myoplabelar{W0}{2.2722in}{1.5014in}
\myoplabelar{T0}{1.2211in}{1.5014in}
\myoplabelar{L0}{1.2211in}{1.5014in}

\myoplabelar{W2}{2.2722in}{1.5014in}
\myoplabelar{T2}{1.2211in}{1.5014in}
\myoplabelar{L2}{1.2211in}{1.5014in}

\myoplabelar{W4}{2.2722in}{1.5014in}
\myoplabelar{T4}{1.2211in}{1.5014in}
\myoplabelar{L4}{1.2211in}{1.5014in}

\myoplabelar{W6}{2.2722in}{1.5014in}
\myoplabelar{T6}{1.2211in}{1.5014in}
\myoplabelar{L6}{1.2211in}{1.5014in}

\end{tikzpicture}
\end{center}
\caption{Contours of spanwise vorticity $\omega_y$ (a, d, g and j),
thrust density $\dt$ (b, e, h and k) and 
lift density $\dl$ (c, d, i and l) of case B090 at four time instants. 
See the corresponding animations in the additional material.
\label{flowvis_detail}}
\end{figure}
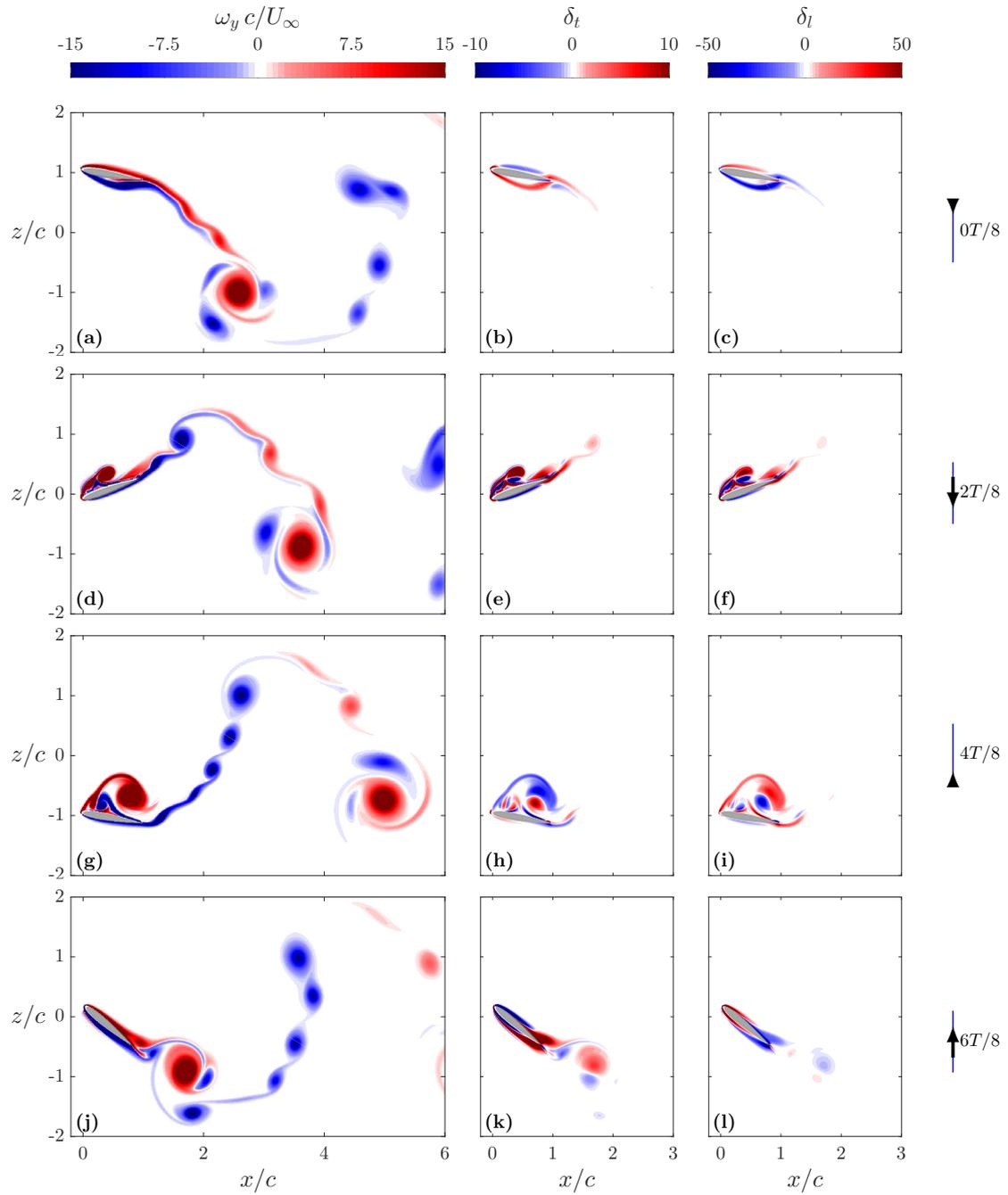
\def\mf{\ref{flowvis_detail}}
%-----------------------------------------flow-visualization-force-density------
In order to obtain a better understanding, we continue the analysis of case B090
by comparing the spanwise vorticity, $\omega_y$, and the force density fields 
of the contribution of vorticity within the flow
at four equispaced time instants during one period, shown in figure \mf{}.
%
%--------------------------------------------------------force-density-decay----
Both thrust \dt{} and lift \dl{} densities (figures \mf{}b, e, h, k and \mf{}c,
f, i, l, respectively) decay fast with the distance to the airfoil for any time
instant, as previously observed by other authors 
\citep{chang1992potential,martin2015vortex}.
This occurs because the force density is the projection of the Lamb's vector on 
the gradient of the auxiliary potentials \phix{} and \phiz{}, which decay 
quadratically with the distance to the airfoil \citep{martin2015vortex}.
%--------------------------------------------------------------vorticity-evo----
The evolution of the spanwise vorticity (figures \mf{}a, d, g and j) shows how
the LEV is created during the downstroke (figure \mf{}d) and shed into the wake 
approximately in the transition from downstroke to upstroke (figure \mf{}g).
At that time, the contribution of the LEV to the thrust changes sign, while its
contribution to the lift remains positive for longer times.
After being shed (figure \mf{}j), the LEV is advected into the wake (figures 
\mf{}a, d and g), and its contribution to the aerodynamic forces becomes 
negligible.
%--------------------------------------------------------------density-evo------
The small influence on the aerodynamic forces from the vortices in the wake is 
consistent with the results of \citet{moriche:2016}, where 2D and 3D
configurations of infinite aspect ratio wings were found to yield very similar
lift and thrust.
%_______________________________________________________________________________

%---------------------------------------------------------vortex-pos/neg-contr--
It is possible to observe in figure \mf{} that the contribution of the LEV to
the thrust and lift is partially positive and negative.
This is an inherent property of any vortex, as indicated by 
\citet{chang1992potential}: the center of a vortex is characterized by a change
of direction of the Lamb's vector while the gradient of the auxiliary potentials
is locally smooth, so a line where the force density changes its sign must pass
through the center of the vortex.
Which part of the vortex (positive and negative contribution to the force) 
dominates, depends on the vorticity field and on the gradients of the auxiliary
potentials, which are determined only by the geometry of the airfoil.
This is an interesting fact, which could be used to guide an optimisation of the
airfoil shape.  
%-------------------------------------------------------------------------------

%---------------------------------------------------------Kutta-Joukowsky-dev---
It is clear from the previous discussion that the contribution of the vorticity 
within the flow is the dominant contribution to the total aerodynamic force.
Therefore it is of great interest to model this part of the force appropriately.
%---------------------------------------------KJ-quasi-steady-==>-steady?-------
In steady-state aerodynamics and in several reduced order models in the
literature (e.g, \citealp{pesavento2004falling, andersen2005unsteady,
taha2014state}), it is common to use the Kutta-Joukowsky theorem to model the
aerodynamic force due to the vorticity within the flow, 
\begin{equation}\label{eq:kuttajoukowsky}
\vec{F}_{KJ} = \rho \left( \vec{U} \times \Gamma\,\vec{e}_y \right)\text{,}
\end{equation}
where the subscript $KJ$ denotes Kutta-Joukowsky estimation, $\Gamma$ is the
circulation around the airfoil, $\vec{U}$ is the effective  velocity seen by
the airfoil and $\vec{e}_y$ is the unitary vector in the $y$ direction.
The force estimated by Kutta-Joukowsky theorem is perpendicular to the incoming
effective velocity, which can be estimated for a heaving and pitching airfoil as
$\vec{U} = U_\infty\,\vec{e}_x - \dot{h}\vec{e}_z$, as described in 
\citet{pesavento2004falling}, \citet{andersen2005unsteady} and
\citet{taha2014state}.
%_______________________________________________________________________________

%-------------------------------------------------------------------------------
\begin{figure}
\begin{center}
\begin{tikzpicture}
\node(A) at (0., 0.) {\ig[width=0.4\textwidth]{fig7.eps}};
\end{tikzpicture}
\caption{Sketch of the deviation angle $\beta$ and effective angle of attack 
$\alpha_e$.
\label{outline_beta}}
\end{center}
\end{figure}
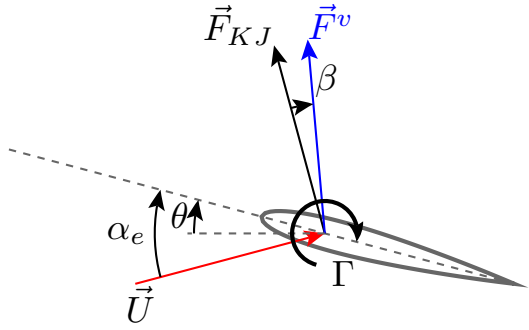
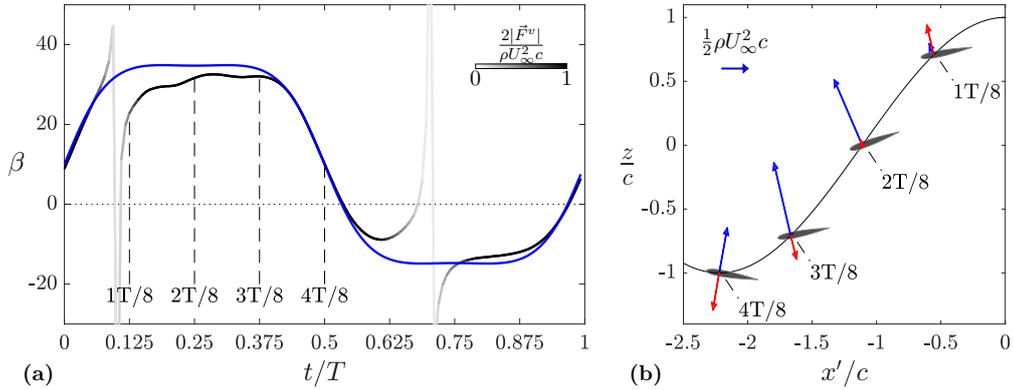
\begin{figure}
\begin{center}
\begin{tikzpicture}
\coordinate(O) at (0cm, 0cm);
\node(B1003) at (O)                 {\ig[scale=0.5]{fig8a.eps}};
\node(F1003) at (B1003.east) [right]{\ig[scale=0.5]{fig8b.eps}};
\setcounter{myplotslabel}{1}
\mylabel{B1003}{0.3in}{0.2in}
\mylabel{F1003}{0.3in}{0.2in}
\coordinate(AXX) at ([xshift= 6.3cm,yshift=-1.10cm]B1003.north west);
\coordinate(BXX) at ([xshift= 1.2cm,yshift= 0.00cm]AXX);
\shade[left color=white, right color=black](AXX) rectangle ([yshift=0.06cm]BXX);
\draw[color=black, line width=0.1mm ](AXX) rectangle ([yshift=0.06cm]BXX);
\node(LAXX) at ([xshift=-0.0cm,yshift=-0.03cm]AXX.south)[below,scale=0.75]{$0$};
\node(LBXX) at ([xshift=-0.0cm,yshift=-0.03cm]BXX.south)[below,scale=0.75]{$1$};
\node(LLXX) at ([xshift=2.2cm,yshift=3mm]LBXX)[above,scale=0.75]{$\frac{1}{2}\rho U_\infty^2 c$};
\node(L2XX) at ([xshift=0.6cm,yshift=0.2cm]LAXX)[above,scale=0.75]%
{$\frac{2|\vec{F}^v|}{\rho U_\infty^2 c}$};
\end{tikzpicture}
\end{center}
\caption{a) Evolution of angle $\beta$ (in degrees) of case B090 during one 
period. The curves represented are the angle $\beta$ in gray scale to indicate 
the modulus of the contribution of vorticity within the flow and the effective 
angle of attack
(\protect\lineSymbolRGB[solid]{0}{0}{1}{0.012in}{none}{0in}). b) Vectors of 
$\vec{F}^v$ (\protect\lineSymbolRGB[solid]{0}{0}{1}{0.012in}{none}{0in}) and
$\vec{F}^m$ (\protect\lineSymbolRGB[solid]{1}{0}{0}{0.012in}{none}{0in})
at four time instants during the downstroke of case B090.
\label{normaldev}}
\end{figure}

%------------------------------------------------------------------beta---------
In order to compare $\vec{F}_{KJ}$ and $\vec{F}^v$, we define the angle $\beta$
formed by these two vectors, as sketched in figure \ref{outline_beta}. 
The evolution of this angle $\beta$ for case B090 during one period is shown in
figure \ref{normaldev}a, where the curve representing $\beta$ is coloured with
a grey scale proportional to $|\vec{F}^v|$.
This is done to avoid confusion when the force tends to zero, situation where
the angle $\beta$ is ill-defined.
The evolution of the effective angle of attack $\alpha_e$ has been also included
in the figure, and it can be clearly seen that the angle $\beta$ tends to 
$\alpha_e$, except for specific time instants where the force is small.
This means that $\vec{F}^v$ is not oriented in the direction suggested by the
Kutta-Joukowsky theorem (namely, normal to the incoming effective velocity), but
$\vec{F}^v$ is approximately perpendicular to the chord of the airfoil.
This observation is consistent with the empirical model of
\cite{dickinson1999wing}, which results in aerodynamic forces essentially normal
to the wing. 
The tendency to a chord-normal orientation of $\vec{F}^v$ is observed for all 
the periodic cases of our database.  
%_______________________________________________________________________________

%-------------------------------------------------------------------------------
Figure \ref{normaldev}b shows a sketch of $\vec{F}^v$ and $\vec{F}^m$ acting on 
the airfoil at four equispaced time instants. 
In the figure the airfoil trajectory is represented as seen by an observer 
travelling with the free stream, so the horizontal coordinate is defined as
$x' = x - U_\infty t$.
Note that the approximate chord-normal orientation of $\vec{F}^v$ occurs when
the forces are large enough, which is consistent with a well developed LEV
moving the suction peak from the leading edge of the airfoil to the upper
surface of the airfoil.
Also, figure \ref{normaldev}a shows that there is a small deviation of
$\vec{F}^v$ with respect to the normal, of the order of $\pm 5\degree$, which
could result in a small component of $\vec{F}^v$ in the direction of the chord.  
Interestingly, $\vec{F}^m$ is also approximately perpendicular to the airfoil
chord.
Indeed, equations \eqref{eq:bm} show that this is strictly true for a flat plate
pitching around $x_p = c/2$.
%_______________________________________________________________________________

%----------------------------------------------------circulation-model-wang-----
From the point of view of modelling, the results of figure \ref{normaldev}
suggest that $\vec{F}^v$ is a force roughly perpendicular to the airfoil chord. 
In order to estimate its modulus, we follow previous works and use the
Kutta-Joukowsky theorem 
\begin{equation}\label{eq:rhogammau}
|\vec{F}^v|=\rho\Gamma |\vec{U}| \text{,}
\end{equation}
estimating the circulation of the airfoil as in \cite{pesavento2004falling}, 
\begin{equation}\label{eq:wangmodel}
\Gamma = \frac{1}{2}G_T\,c \left| \vec{U}\right| \sin(2\alpha_e)
       + \frac{1}{2}G_R\,c^2 \dot{\theta} \text{.}
\end{equation}
In this expression, $G_T$ and $G_R$ are the two free parameters of the model.
Note that \eqref{eq:wangmodel} includes information of the amplitudes and
frequency of the pitching and heaving motion of the airfoil in $\vec{U},
\alpha_e$ and $\theta$. 

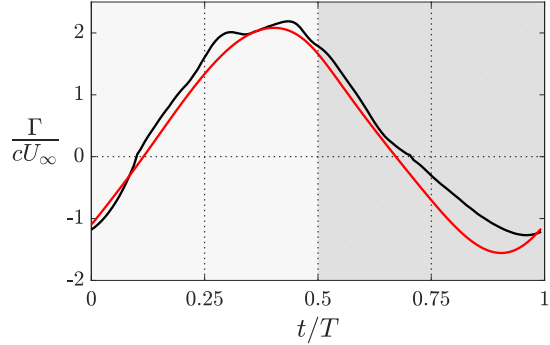
\begin{figure}
\begin{center}
\begin{tikzpicture}
\coordinate(O) at (0, 0);
\node(B) at (O) {\ig[scale=0.5]{fig9.eps}};
\end{tikzpicture}
\caption{Circulation of the airfoil for case B090. The curves represented 
correspond to the values obtained from the DNS 
(\lineSymbolRGB[solid]{0}{0}{0}{0.012in}{none}{0in}) and the model best fit 
(\lineSymbolRGB[solid]{1}{0}{0}{0.012in}{none}{0in}) with $G_T = 1.65$ and
$G_R = 3.73$.
The downstroke (upstroke) is indicated by a light (dark) grey background.
\label{circulation}}
\end{center}
\end{figure}

Figure \ref{circulation} shows the comparison of the circulation obtained from
the DNS (using the computed $\vec{F}^v$ and equation \ref{eq:rhogammau}) and the
circulation obtained from \eqref{eq:wangmodel}.
For the latter, the constants $G_T=1.65$ and $G_R=3.75$ are obtained from a
least square fit to the circulation from the DNS. 
It can be seen that the agreement is very good, except for the last part of the
upstroke ($t/T>0.75$), where the model slightly overpredicts the circulation.
%_______________________________________________________________________________

%-------------------------------------------------------------------------------
%-------------------------------------------------------------------------------

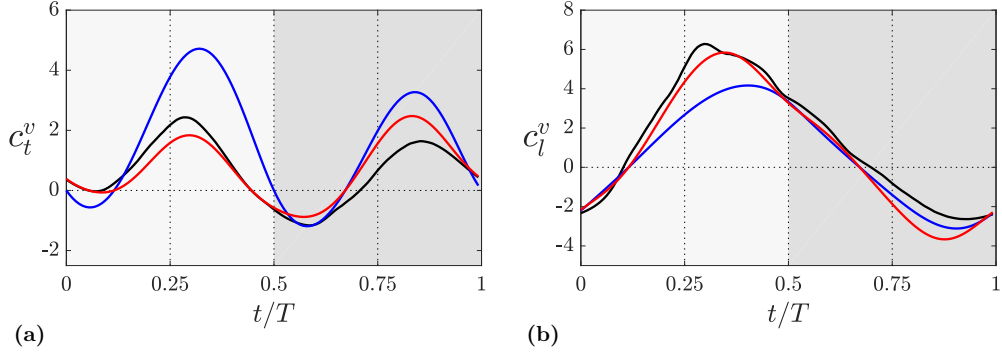
\begin{figure}
\begin{center}
\begin{tikzpicture}
\coordinate(O) at (0, 0);
\node(A) at (O)             {\ig[scale=0.5]{fig10a.eps}};
\node(B) at (A.east) [right]{\ig[scale=0.5]{fig10b.eps}};
\setcounter{myplotslabel}{1}
\mylabel{A}{0.3in}{0.1in}
\mylabel{B}{0.3in}{0.1in}
\end{tikzpicture}
\caption{%\protect\input{figtab/\tbm/\tbm_21003_data/\tbm_21003.glb}
Evolution of a) \ctv and b) \clv of case B090 during one period. The curves 
represented correspond to the values obtained from the DNS 
(\lineSymbolRGB[solid]{0}{0}{0}{0.012in}{none}{0in}), the Kutta-Joukowsky 
estimation (\lineSymbolRGB[solid]{0}{0}{1}{0.012in}{none}{0in}) and the 
chord-normal estimation 
(\lineSymbolRGB[solid]{1}{0}{0}{0.012in}{none}{0in}). The circulation in
both estimations is given by the model from Pesavento and Wang with $G_T=1.65$
and $G_R=3.73$.
The downstroke (upstroke) is indicated by a light (dark) grey background.
\label{forcetilted_detail}}
\end{center}
\end{figure}

%----------------------------------------------------chord-normal-estimation----
Based on the observations made in this section, we propose modelling $\vec{F}^v$
as a force oriented normal to the chord, whose modulus is calculated with
equations \eqref{eq:rhogammau} and \eqref{eq:wangmodel}. 
%-------------------------------------------------------------comments-on-figure
Figure \ref{forcetilted_detail} shows the comparison between this chord-normal
model, the Kutta-Joukowsky prediction (i.e., same modulus but perpendicular to
$\vec{U}$) and the results of the DNS. 
The results show that peak values of thrust (figure \ref{forcetilted_detail}a)
are well predicted by the chord-normal force, but overestimated by 
Kutta-Joukowsky prediction.
Regarding the lift, figure \ref{forcetilted_detail}b shows that \clv{} is well
reproduced by the chord-normal estimation, with a visible deviation 
corresponding to the already mentioned misrepresentation of $\Gamma$ near the
end of the upstroke. 
%_______________________________________________________________________________

%===============================================================================
%===============================================================================
%===============================================================================
\section{Effects of mean pitch angle and phase shift in the force decomposition
and modelling}
\label{sec:ext}

%-------------------------------------------------------------------------------
In this section we extend the analysis performed in the previous section on case
B090 to a subset of cases from the database.
The objective of extending the analysis is to see how the different
contributions to the total aerodynamic force are influenced by the motion 
parameters \thm{} and \pphi{}.
%-------------------------------------differences-volume-component-mean-pitch---
As before, we select the cases A090, C090, B070 and B110 and use the results of
case B090 as a reference.
%_______________________________________________________________________________

%-------------------------------------------------------------------------------
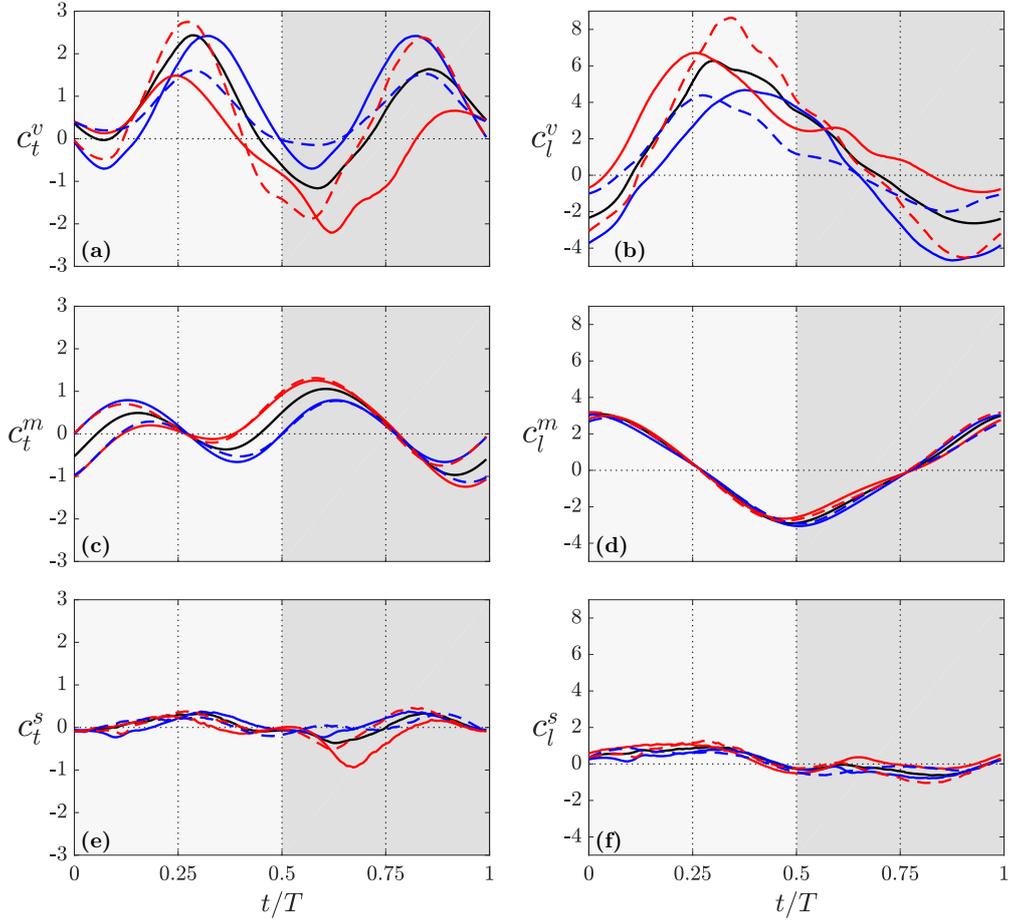
\begin{figure}
\begin{center}
\begin{tikzpicture}
\coordinate(O) at (0cm, 0cm);
% THRUST
\node[](T2) at (O)              {\ig[scale=0.5]{fig11a.eps}};
\node[](T3) at (T2.south)[below]{\ig[scale=0.5]{fig11b.eps}};
\node[](T4) at (T3.south)[below]{\ig[scale=0.5]{fig11c.eps}};
% LIFT                                                
\node[](L2) at (T2.east) [right]{\ig[scale=0.5]{fig11d.eps}};
\node[](L3) at (L2.south)[below]{\ig[scale=0.5]{fig11e.eps}};
\node[](L4) at (L3.south)[below]{\ig[scale=0.5]{fig11f.eps}};
\setcounter{myplotslabel}{1}
\myoplabelar{T2}{2.250in}{1.4332in}\myoplabelar{L2}{2.1572in}{1.4332in}
\myoplabelar{T3}{2.250in}{1.4332in}\myoplabelar{L3}{2.250in}{1.4332in}
\myoplabelar{T4}{2.250in}{1.4332in}\myoplabelar{L4}{2.250in}{1.4332in}
\end{tikzpicture}
\end{center}
\caption{%\protect\input{figtab/\tbm/\tbm_2000_data/\tbm_2000.glb}
Contribution of a) vorticity within the flow, c) body motion and e)
surface vorticity to the total aerodynamic thrust. The lift contributions are 
shown in b), d) and f), respectively. Four cases with different motion 
parameters are represented:
case B090 (\lineSymbolRGB[solid]{0}{0}{0}{0.012in}{none}{0in}),
case A090 (\lineSymbolRGB[solid]{0}{0}{1}{0.012in}{none}{0in}),
case C090 (\lineSymbolRGB[solid]{1}{0}{0}{0.012in}{none}{0in}),
case B070 (\lineSymbolRGB[dashed]{0}{0}{1}{0.012in}{none}{0in}) and
case B110 (\lineSymbolRGB[dashed]{1}{0}{0}{0.012in}{none}{0in}).
The downstroke (upstroke) is indicated by a light (dark) grey background.
\label{forcesdecomp_focus}}
\end{figure}
\def\mf{\ref{forcesdecomp_focus}} 
%-------------------------------------------------------------------------------
Figure \mf{} shows the contributions to the lift and thrust coefficients from
the vorticity within the flow, body motion and surface vorticity for the
selected cases. 
Overall, figure \mf{}a and b show that increasing $\theta_m$ increases $\clv{}$
and reduces $\ctv{}$. This is because increasing $\theta_m$ increases the
effective angle of attack (figure \ref{forcesdecomp_focus_TOTAL}c), increasing
the modulus of $\vec{F}^v$, and tilting the resultant backwards.
Hence, reducing thrust and increasing lift.
Also, the LEV detaches from the airfoil earlier as $\theta_m$ increases,
shifting the peaks in lift during the downstroke to earlier times (see figure
\mf{}b).
As mentioned earlier, this has a small effect on the thrust, which is strongly
influenced by the orientation of the airfoil.
During the upstroke, increasing $\theta_m$ reduces the (negative) effective
angle of attack.
This results in lower forces, although better oriented for thrust.
For C090, at the beginning of the upstroke the pitch angle is large enough so
that the airfoil is still producing positive lift, and since it is tilted
backwards, drag  (see $\theta$ and $\alpha_e$ in figure
\ref{forcesdecomp_focus_TOTAL}c and d, respectively).
%_______________________________________________________________________________

%--------------------------------------differences-volume-component-phase-------
As with the total forces, the variation of phase shift produces a different
effect, affecting mainly the amplitude of the peak values of $\vec{F}^v$.
An advance of the pitching motion (case B110) increases the amplitude of the
fluctuations of both thrust and lift, whereas lagging the pitching motion (case
B070) results in a reduction of these amplitudes.  
There is also an effect of the phase shift in the position of the peaks of
$c_l^v$ during the downstroke, although this effect is not appreciable in
$c_l^v$ during the upstroke, nor in the peaks of $c_t^v$. 
%-------------------------------------------------------------------------------

%--------------------------------------differences-added-component--------------
Figures \mf{}c and d show the evolution of \cta{} and \cla{}, respectively, for 
the selected cases.
For the motion under study, the heaving acceleration $\ddot{h}(t)$ dominates the
contribution of body motion to the lift, independently of the parameters \thm{}
and \pphi{} (figure \mf{}d).
Concerning the thrust, both the heaving acceleration $\ddot{h}(t)$ and the
projected area of the airfoil perpendicular to the streamwise direction 
influence \cta{}.
Therefore, \cta{} vanishes at points where the heaving acceleration 
$\ddot{h}(t)$ is zero ($t/T = 0.25, 0.75$). At points where the heaving 
acceleration $\ddot{h}(t)$ is maximum ($t/T=0,0.5$), the value of \cta{}
depends on the area of the airfoil projected perpendicular to the streamwise 
direction.
This area  is related to the value of the pitch angle $\theta$ (figure 
\ref{forcesdecomp_focus_TOTAL}e).
At the beginning of the downstroke ($t/T=0$), cases A090 and B110 have a pitch
angle $\theta$ of $0\degree$, so \cta{} is roughly zero (figure \mf{}c).
At this time instant, \cta{} increases with the instantanous pitch angle, so
case B090 generates more thrust than cases A090 and B110 and less than cases
C090 and B070.
The same analysis holds at the beginning of the upstroke ($t/T=0.5$), except
that, in this part of the period, cases with different phase shift (B070 and 
B110) interchange their role.
%-------------------------------------------------------------------------------

%--------------------------------------differences-surface-vorticity------------
To conclude the analysis of the different contributions to the total aerodynamic
force, figures \ref{forcesdecomp_focus}e and f show the evolution of the surface
vorticity contribution to the total aerodynamic thrust and lift, respectively.
It can be seen that the effects that surface vorticity has on the forces is 
small compared to the other contributions for both thrust and lift, except for a
peak of viscous drag in the upstroke of case C090.
A possible explanation is related to the fact that the effective angle of attack
during the upstroke is close to zero, as shown in figure
\ref{forcesdecomp_focus_TOTAL}d, unlike the other cases considered.
As a consequence, during the first half of the upstroke, the boundary layer in 
the lower surface is attached and thinner than in the other cases, with no
appreciable LEV in the lower surface of the airfoil.
This has been observed in visual inspection of the corresponding flow fields
(not shown here).
Hence, the skin friction in the lower surface of the airfoil is larger for case
C090.
%_______________________________________________________________________________

\begin{table}
\begin{tabular*}{\textwidth}{@{\extracolsep{\fill}}lcccccc}
Case &            \thm  &             \pphi & $G_T$&$G_R$ & \multicolumn{2}{c}{$\|\Gamma_{DNS}-\Gamma\|_2$} \\
     & & & & & Specific & Fixed  \\
\hline 
B090&$\om0.00\degree$ & $\om90.00\degree$ & $1.65$ & $3.73$ & $0.26$ & $0.32$  \\
A090&$  10.00\degree$ & $\om90.00\degree$ & $1.44$ & $4.48$ & $0.13$ & $0.43$  \\
C090&$  20.00\degree$ & $\om90.00\degree$ & $1.95$ & $2.14$ & $0.42$ & $0.50$  \\
B070&$  10.00\degree$ & $\om70.00\degree$ & $1.65$ & $2.46$ & $0.14$ & $0.20$  \\
B110&$  10.00\degree$ & $  110.00\degree$ & $1.34$ & $4.58$ & $0.35$ & $0.50$  
\end{tabular*}
\caption{Coefficients of \citet{pesavento2004falling} model for circulation 
(equation \eqref{eq:wangmodel}) obtained from the best fit with the data from 
the DNS for the selected cases. The errors shown correspond to the circulation
obtained with the coefficients obtained specifically for each case 
(next-to-last) column and fixed coefficients $G_T=1.85$ and $G_R=\pi$ (last 
column).
\label{tab:wangcoeff}}
\end{table}
%--------------------------------------------Wang-focus-cases-------------------
After describing the evolution of the different contributions to the total
aerodynamic force, we proceed to check the capability of the chord-normal
model proposed in the previous section to predict the contribution of the
vorticity within the flow to the total aerodynamic force of the subset of cases
A090, C090, B070 and B110.
The first question is which values of $G_T$ and $G_R$ should be used in equation 
\eqref{eq:wangmodel}. 
In principle, one could obtain ``specific'' values for $G_T$ and $G_R$ for each
case, repeating the same least square fitting process described in section
\ref{sec:ref} but for cases A090, C090, B070 and B110. 
These specific coefficients, as well as the corresponding L2 norm of the
difference between the model and DNS circulation, are shown in table
\ref{tab:wangcoeff}. 
It should be noted that, even if the variability of the coefficients with
$\varphi$ and $\theta_m$ is not small (around $\pm 19$\% for $G_T$ and
$\pm 35$\% for $G_R$), the assumption in \citet{pesavento2004falling} is that
the coefficients $G_T$ and $G_R$ depend on the geometry, not on the kinematics.
Indeed, the kinematic of the airfoil enters in equation \eqref{eq:wangmodel}
through $\alpha_e$ and $\theta$. 
%_______________________________________________________________________________

%-------------------------------------------------------------------------------
To check the validity of this assumption, we have also computed ``fixed'' values
of these constants, choosing $G_R=\pi$ and fitting $G_T$ for all the periodic 
cases in table \ref{tab:clct_all} using a least square method. 
The rationale for choosing $G_R=\pi$ comes from potential theory, which results
in
\begin{equation}\label{eq:gammarot}
\Gamma_{ROT} = \pi \,c^2 \dot{\theta}\left(\frac{3}{4} - \frac{x_p}{c}\right) 
\end{equation}
for a thin airfoil in piching motion (this can be derived from the expressions
appearing in section 6.2 of \citealp{fung2002introduction}).
The resulting $G_T=1.85$, which is approximately the mean value of the 
``specific'' values obtained for $G_T$, yields errors in the circulation
$\Gamma$ of the same order of magnitude (see last column in table
\ref{tab:wangcoeff}).  
Note also that from the point of view of modelling, ``fixed'' values of $G_T$
and $G_R$ are more interesting than ``specific'' values for these constants,
since the former can be used to predict the circulation without having to run a
DNS or a experiment. 
We have tried to fit the values of $G_T$ and $G_R$ in slightly different ways
(for instance, $G_T=1$ and fit $G_R$), but in all cases the $L_2$ norms of the
difference in the circulation were similar as those presented in table
\ref{tab:wangcoeff}. 
From that point of view, it could be argued that the results presented below
are robust with respect to the values of these two parameters. 
%_______________________________________________________________________________

%-------------------------------------------------------------------------------
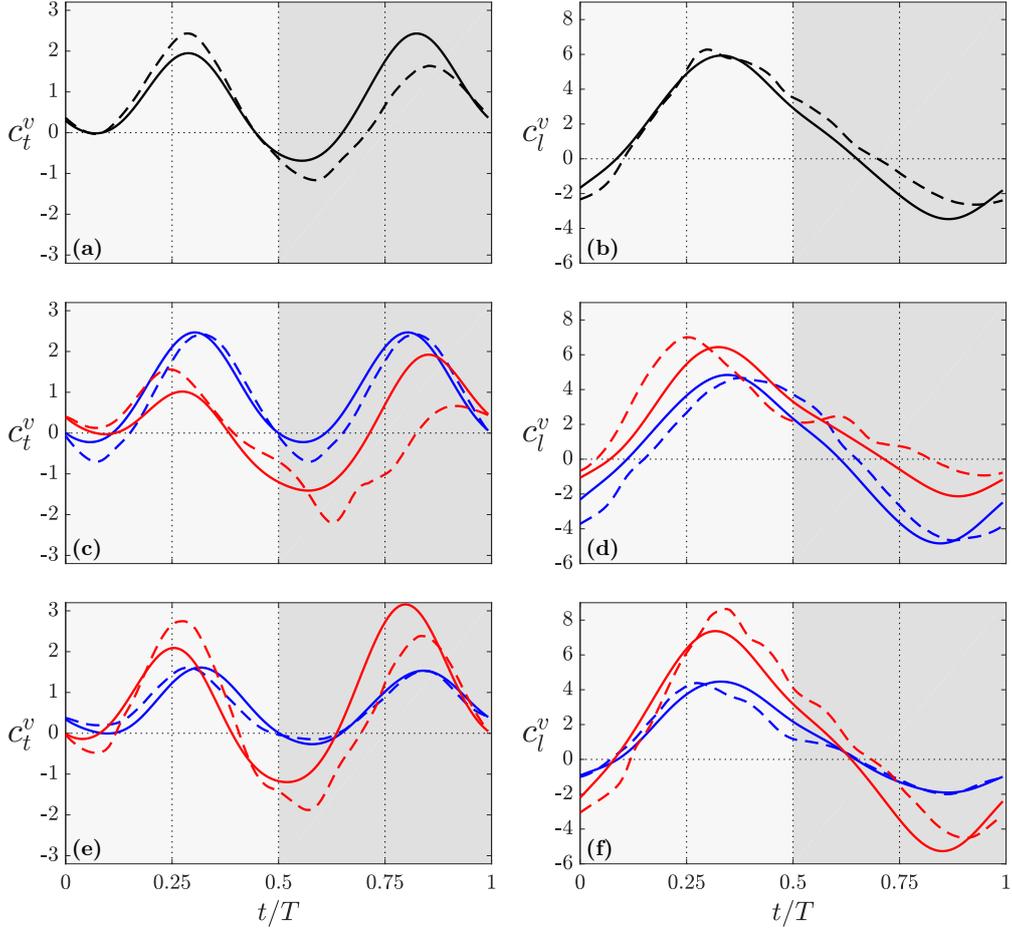
\begin{figure}
\begin{center}
\begin{tikzpicture}
\coordinate(O) at (0cm, 0cm);
% THRUST
\node[](T1) at (O)              {\ig[scale=0.5]{fig12a.eps}};
\node[](T2) at (T1.south)[below]{\ig[scale=0.5]{fig12b.eps}};
\node[](T3) at (T2.south)[below]{\ig[scale=0.5]{fig12c.eps}};
                                                      
\node[](L1) at (T1.east) [right]{\ig[scale=0.5]{fig12d.eps}};
\node[](L2) at (T2.east) [right]{\ig[scale=0.5]{fig12e.eps}};
\node[](L3) at (L2.south)[below]{\ig[scale=0.5]{fig12f.eps}};
\setcounter{myplotslabel}{1}
\myoplabelar{T1}{2.3072in}{1.4641in}\myoplabelar{L1}{2.3072in}{1.4641in}
\myoplabelar{T2}{2.3072in}{1.4641in}\myoplabelar{L2}{2.3072in}{1.4641in}
\myoplabelar{T3}{2.3072in}{1.4641in}\myoplabelar{L3}{2.3072in}{1.4641in}
\end{tikzpicture}

\caption{%\protect\input{figtab/\tbm/\tbm_10000_data/\tbm_10000.glb}
Chord-normal estimation of the contribution of vorticity within the flow
for the selected cases. The circulation is given by the model (equation 
\eqref{eq:wangmodel}) with coefficients $G_T = 1.85$ and $G_R = \pi$.
a) Thrust and b) lift of case B090
(DNS  \lineSymbolRGB[dashed]{0}{0}{0}{0.012in}{none}{0in} and 
model \lineSymbolRGB[solid]{0}{0}{0}{0.012in}{none}{0in}).
c) Thrust and d) lift of cases A090
(DNS  \lineSymbolRGB[dashed]{0}{0}{1}{0.012in}{none}{0in} and 
model \lineSymbolRGB[solid]{0}{0}{1}{0.012in}{none}{0in}) and C090
(DNS  \lineSymbolRGB[dashed]{1}{0}{0}{0.012in}{none}{0in} and 
model \lineSymbolRGB[solid]{1}{0}{0}{0.012in}{none}{0in}). 
e) Thrust and f) lift of cases B070 
(DNS  \lineSymbolRGB[dashed]{0}{0}{1}{0.012in}{none}{0in} and 
model \lineSymbolRGB[solid]{0}{0}{1}{0.012in}{none}{0in}) and B110 
(DNS  \lineSymbolRGB[dashed]{1}{0}{0}{0.012in}{none}{0in} and
model \lineSymbolRGB[solid]{1}{0}{0}{0.012in}{none}{0in}). 
The downstroke (upstroke) is indicated by a light (dark) grey background.
\label{forcetilted_focus}}
\end{center}
\end{figure}
%-------------------------------------------------comment-on-figure-------------

\def\mf{\ref{forcetilted_focus}}
%----------------------------------------------lift thrust estimation-----------
Figure \mf{} shows the goodness of the chord-normal model ($\vec{F}^v$
perpendicular to the chord, modulus of $\vec{F}^v$ given by equations
\eqref{eq:wangmodel} and \eqref{eq:rhogammau}, $G_T=1.85$ and $G_R = \pi$) to
predict \ctv{} and \clv{} for the selected cases. 
Results for case B090 are shown in figures \mf{}a and b, showing similar results
to those obtained when using specific coefficients (figure
\ref{forcetilted_detail}). 
This is consistent with the values of the $L_2$ norm shown in table
\ref{tab:wangcoeff}. 
Figures \mf{}c and d show the evolution of $\ctv$ and $\clv$, respectively, for 
cases A090 and C090 (varying the mean pitch angle).
Both \ctv{} and \clv{} are properly predicted for case A090, but the differences
between the chord-normal estimation and the results obtained from the DNS are 
larger for case C090.
These differences consist on both a small time shift and a change in the peak
values. 
Figures \mf{}e and f presents the comparison of model and DNS for B070 and B110
(varying the phase shift), showing a better prediction for case B070 than for
case B110. 
Overall, the results in figure \mf{} suggest that the chord normal model for
the vorticity within the flow works better for moderate $\theta_m$ and
$\varphi \lesssim 90 \degree$.
%_______________________________________________________________________________

%===============================================================================
%===============================================================================
%===============================================================================
\section{Prediction of total aerodynamic forces}
\label{sec:tot}

%-------------------------------------------------------------------------------
In this last section we estimate the total aerodynamic force taking into account
the results presented throughout the manuscript. 
We discuss two different models, namely  M1 and M2. 
While M1 introduces features based on observations made through
this work, M2 is selected for comparison, based on previous works
\citep{andersen2005unsteady,pesavento2004falling}.
% 
% REBUTTAL BEG ----------------------------------------------------------------
Both models consider inertial effects originated from the motion of the airfoil
and forces produced by the vorticity within the flow.
In M1, the inertial force is the body motion term of \citet{chang1992potential},  
$\vec{F}^m$, and it is computed using \eqref{eq:force_decomp} and the auxiliary
potential given in figure \ref{fig:phiatfoil}. 
The force due to the vorticity within the flow is estimated as discussed in the
previous section: $\vec{F}^v$ is normal to the airfoil chord, with a modulus
given by \eqref{eq:rhogammau} and \eqref{eq:wangmodel}.
%_______________________________________________________________________________

%-------------------------------------------------------------------------------
On the other hand, M2 is the model proposed by \citet{andersen2005unsteady}, but
without the viscous terms. The effect of the body motion is estimated using the
added mass expressions of \citet{sedov1965two} for a flat plate hinged at $c/4$.
The contribution from the vorticity within the flow is estimated as a force
perpendicular to the effective velocity $\vec{U}$, with modulus given by
expressions \eqref{eq:rhogammau} and \eqref{eq:wangmodel}.
%_______________________________________________________________________________

%-------------------------------------------------------------------------------
Note that in order to minimize the differences between M1 and M2, the
coefficients in \eqref{eq:wangmodel} are $C_T=1.85$ and $C_R=\pi$ for both
models. 
Still, there are two important differences between M1 and M2. 
First, the nature of the contribution from the inertial forces, which in M1 can
yield a net force (i.e, $\ctm^m$ and $\clm^m$ are not necessarily zero) while
in M2 does not yield a net force. 
Second, the orientation of the contribution from the vorticity within the flow,
which is one of the main observations of section \ref{sec:ref}. 
%_______________________________________________________________________________

%-------------------------------------------------------------------------------
% script 1: ppot_modeltotal_a1_all_pre
% script 2: ppot_modeltotal_a1_all
\begin{table}
%\begin{tabular*}{\textwidth}{@{\extracolsep{\fill}}|l|cc|cc|cc|cc|} 
\begin{tabular*}{\textwidth}{@{\extracolsep{\fill}}|l|cc|cc|cc|cc|} 
     &    \multicolumn{2}{c|}{$\overline{c}_{t,M\#}-\overline{c}_{t,DNS}$}%$\overline{\ct - \ct_{DNS}}$ } 
     &    \multicolumn{2}{c|}{$\|c_{t,M\#}-c_{t,DNS}\|_2$ }%L_2\left(\ct - \ct_{DNS}\right)$} 
     &    \multicolumn{2}{c|}{$\overline{c}_{l,M\#}-\overline{c}_{l,DNS}$ }%$\overline{\cl} - \overline{\cl}_{DNS}$ } 
     &    \multicolumn{2}{c|}{$\|c_{l,M\#}-c_{l,DNS}\|_2$ } \\ %L_2\left(\cl - \cl_{DNS}\right)$} \\
\hline
Case &           M1    &           M2 &           M1 &          M2  &              M1 &              M2 &           M1 &           M2  \\
A030 & $ \im 0.226\im$ & $\im 1.109\im$ & $\im     0.444\im$ & $\im     1.168\im$ & $ \im 0.004\im$ & $ \im 0.004\im$ & $\im     1.426\im$ & $\im     2.428\im$  \\
A050 & $ \im 0.106\im$ & $\im 0.785\im$ & $\im     0.402\im$ & $\im     0.994\im$ & $ \im 0.000\im$ & $ \im 0.000\im$ & $\im     0.689\im$ & $\im     1.793\im$  \\
A070 & $    -0.066\im$ & $\im 0.513\im$ & $\im     0.131\im$ & $\im     0.808\im$ & $    -0.000\im$ & $    -0.000\im$ & $\im     0.268\im$ & $\im     1.004\im$  \\
A090 & $ \im 0.116\im$ & $\im 0.835\im$ & $\im     0.471\im$ & $\im     1.173\im$ & $ \im 0.000\im$ & $    -0.000\im$ & $\im     0.796\im$ & $\im     0.316\im$  \\
A110 & $ \im 0.246\im$ & $\im 1.400\im$ & $\im     0.876\im$ & $\im     1.654\im$ & $    -0.000\im$ & $    -0.000\im$ & $\im     1.245\im$ & $\im     1.040\im$  \\
A130 & $ \im 0.287\im$ & $\im 2.096\im$ & $\im     1.181\im$ & $\im     2.221\im$ & $ \im 0.005\im$ & $ \im 0.005\im$ & $\im     1.896\im$ & $\im     2.235\im$  \\
B030 & $ \im 0.065\im$ & $\im 0.864\im$ & $\im     0.395\im$ & $\im     1.196\im$ & $ \im 0.102\im$ & $    -0.011\im$ & $\im     1.503\im$ & $\im     2.368\im$  \\
B050 & $ \im 0.042\im$ & $\im 0.724\im$ & $\im     0.355\im$ & $\im     1.195\im$ & $    -0.108\im$ & $    -0.043\im$ & $\im     1.038\im$ & $\im     1.953\im$  \\
B070 & $    -0.025\im$ & $\im 0.627\im$ & $\im     0.268\im$ & $\im     1.131\im$ & $ \im 0.010\im$ & $ \im 0.242\im$ & $\im     0.826\im$ & $\im     1.700\im$  \\
B090 & $ \im 0.166\im$ & $\im 0.995\im$ & $\im     0.620\im$ & $\im     1.377\im$ & $ \im 0.557\im$ & $ \im 0.913\im$ & $\im     0.699\im$ & $\im     1.182\im$  \\
B110 & $ \im 0.209\im$ & $\im 1.467\im$ & $\im     0.852\im$ & $\im     1.766\im$ & $ \im 0.597\im$ & $ \im 1.014\im$ & $\im     1.107\im$ & $\im     1.404\im$  \\
B130 & $ \im 0.311\im$ & $\im 2.179\im$ & $\im     1.181\im$ & $\im     2.396\im$ & $ \im 0.805\im$ & $ \im 1.209\im$ & $\im     1.934\im$ & $\im     2.511\im$  \\
C030 & $ \im 0.628\im$ & $\im 1.176\im$ & $\im     1.077\im$ & $\im     1.717\im$ & $ \im 0.984\im$ & $ \im 0.732\im$ & $\im     2.593\im$ & $\im     2.859\im$  \\
C050 & $ \im 0.693\im$ & $\im 1.358\im$ & $\im     0.874\im$ & $\im     1.683\im$ & $ \im 1.345\im$ & $ \im 1.388\im$ & $\im     2.118\im$ & $\im     2.427\im$  \\
C070 & $ \im 0.094\im$ & $\im 0.926\im$ & $\im     0.223\im$ & $\im     1.647\im$ & $ \im 0.307\im$ & $ \im 0.637\im$ & $\im     1.409\im$ & $\im     2.179\im$  \\
C090 & $ \im 0.432\im$ & $\im 1.544\im$ & $\im     1.075\im$ & $\im     2.026\im$ & $ \im 1.032\im$ & $ \im 1.587\im$ & $\im     1.617\im$ & $\im     2.400\im$  \\
C110 & $ \im 0.336\im$ & $\im 1.861\im$ & $\im     1.136\im$ & $\im     2.271\im$ & $ \im 1.240\im$ & $ \im 1.919\im$ & $\im     1.569\im$ & $\im     2.415\im$  \\
C130 & $ \im 0.460\im$ & $\im 2.474\im$ & $\im     1.410\im$ & $\im     2.894\im$ & $ \im 1.654\im$ & $ \im 2.337\im$ & $\im     2.245\im$ & $\im     3.166\im$  
\end{tabular*}
\caption{Mean and $L_2$ norm of the difference between an estimation of the 
thrust and lift and the value obtained from the DNS.
The models presented correspond to the estimation proposed in the present work 
(M1) and an estimation taken from the literature (M2, see
\citealp{andersen2005unsteady}).
\label{tab:total}}
\end{table}
%-------------------------------------------------------------------------------

\begin{figure}
\begin{center}
\begin{tikzpicture}
\coordinate(O) at (0.,0.);
\node(A030) at (O)                {\ig[scale=0.5]{fig13a.eps}};
\node(B030) at (A030.east) [right]{\ig[scale=0.5]{fig13b.eps}};
\node(C030) at (B030.east) [right]{\ig[scale=0.5]{fig13c.eps}};
\node(A050) at (A030.south)[below]{\ig[scale=0.5]{fig13d.eps}};
\node(B050) at (B030.south)[below]{\ig[scale=0.5]{fig13e.eps}};
\node(C050) at (C030.south)[below]{\ig[scale=0.5]{fig13f.eps}};
\node(A070) at (A050.south)[below]{\ig[scale=0.5]{fig13g.eps}};
\node(B070) at (B050.south)[below]{\ig[scale=0.5]{fig13h.eps}};
\node(C070) at (C050.south)[below]{\ig[scale=0.5]{fig13i.eps}};
\node(A090) at (A070.south)[below]{\ig[scale=0.5]{fig13j.eps}};
\node(B090) at (B070.south)[below]{\ig[scale=0.5]{fig13k.eps}};
\node(C090) at (C070.south)[below]{\ig[scale=0.5]{fig13l.eps}};
\node(A110) at (A090.south)[below]{\ig[scale=0.5]{fig13m.eps}};
\node(B110) at (B090.south)[below]{\ig[scale=0.5]{fig13n.eps}};
\node(C110) at (C090.south)[below]{\ig[scale=0.5]{fig13o.eps}};
\node(A130) at (A110.south)[below]{\ig[scale=0.5]{fig13p.eps}};
\node(B130) at (B110.south)[below]{\ig[scale=0.5]{fig13q.eps}};
\node(C130) at (C110.south)[below]{\ig[scale=0.5]{fig13r.eps}};
% labels
\node(TH00) at (A030.north)[above]{$\theta_m= 0\degree$};
\node(TH10) at (B030.north)[above]{$\theta_m=10\degree$};
\node(TH20) at (C030.north)[above]{$\theta_m=20\degree$};
\node(PHI030) at (C030.east)[right]{$\varphi= 30\degree$};
\node(PHI050) at (C050.east)[right]{$\varphi= 50\degree$};
\node(PHI070) at (C070.east)[right]{$\varphi= 70\degree$};
\node(PHI090) at (C090.east)[right]{$\varphi= 90\degree$};
\node(PHI110) at (C110.east)[right]{$\varphi=110\degree$};
\node(PHI130) at (C130.east)[right]{$\varphi=130\degree$};
\setcounter{myplotslabel}{1}
\myoplabelar{A030}{1.5715in}{1.0094in}
\myoplabelar{B030}{1.5715in}{1.0094in}
\myoplabelar{C030}{1.2715in}{1.0094in}
\myoplabelar{A050}{1.5715in}{1.0094in}
\myoplabelar{B050}{1.5715in}{1.0094in}
\myoplabelar{C050}{1.5715in}{1.0094in}
\myoplabelar{A070}{1.5715in}{1.0094in}
\myoplabelar{B070}{1.5715in}{1.0094in}
\myoplabelar{C070}{1.5715in}{1.0094in}
\myoplabelar{A090}{1.5715in}{1.0094in}
\myoplabelar{B090}{1.5715in}{1.0094in}
\myoplabelar{C090}{1.5715in}{1.0094in}
\myoplabelar{A110}{1.5715in}{1.0094in}
\myoplabelar{B110}{1.5715in}{1.0094in}
\myoplabelar{C110}{1.5715in}{1.0094in}
\myoplabelar{A130}{1.5715in}{1.0094in}
\myoplabelar{B130}{1.5715in}{1.0094in}
\myoplabelar{C130}{1.5715in}{1.0094in}
\end{tikzpicture}
\caption{Total thrust obtained from the 
DNS (\lineSymbolRGB[solid]{0}{0}{0}{0.012in}{none}{0in}) together with the 
estimations with M1 (\lineSymbolRGB[solid]{1}{0}{0}{0.012in}{none}{0in}) 
and M2 (\lineSymbolRGB[solid]{0}{0}{1}{0.012in}{none}{0in}).
\label{fig:totct}}
\end{center}
\end{figure}
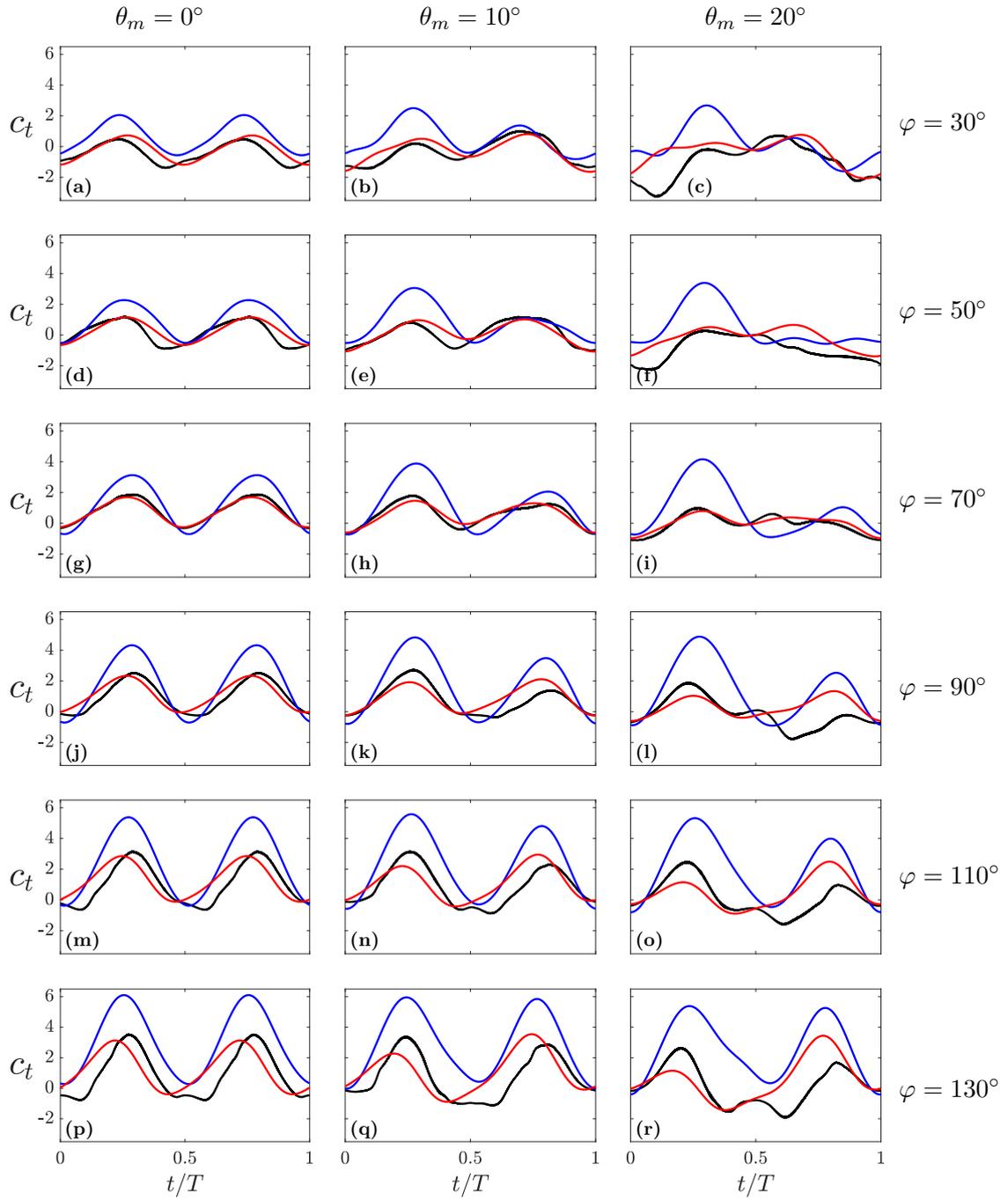

\begin{figure}
\begin{center}
\begin{tikzpicture}
\coordinate(O) at (0.,0.);
\node(A030) at (O)                {\ig[scale=0.5]{fig14a.eps}};
\node(B030) at (A030.east)[right] {\ig[scale=0.5]{fig14b.eps}};
\node(C030) at (B030.east)[right] {\ig[scale=0.5]{fig14c.eps}};
\node(A050) at (A030.south)[below]{\ig[scale=0.5]{fig14d.eps}};
\node(B050) at (B030.south)[below]{\ig[scale=0.5]{fig14e.eps}};
\node(C050) at (C030.south)[below]{\ig[scale=0.5]{fig14f.eps}};
\node(A070) at (A050.south)[below]{\ig[scale=0.5]{fig14g.eps}};
\node(B070) at (B050.south)[below]{\ig[scale=0.5]{fig14h.eps}};
\node(C070) at (C050.south)[below]{\ig[scale=0.5]{fig14i.eps}};
\node(A090) at (A070.south)[below]{\ig[scale=0.5]{fig14j.eps}};
\node(B090) at (B070.south)[below]{\ig[scale=0.5]{fig14k.eps}};
\node(C090) at (C070.south)[below]{\ig[scale=0.5]{fig14l.eps}};
\node(A110) at (A090.south)[below]{\ig[scale=0.5]{fig14m.eps}};
\node(B110) at (B090.south)[below]{\ig[scale=0.5]{fig14n.eps}};
\node(C110) at (C090.south)[below]{\ig[scale=0.5]{fig14o.eps}};
\node(A130) at (A110.south)[below]{\ig[scale=0.5]{fig14p.eps}};
\node(B130) at (B110.south)[below]{\ig[scale=0.5]{fig14q.eps}};
\node(C130) at (C110.south)[below]{\ig[scale=0.5]{fig14r.eps}};
% labels
\node(TH00) at (A030.north)[above]{$\theta_m= 0\degree$};
\node(TH10) at (B030.north)[above]{$\theta_m=10\degree$};
\node(TH20) at (C030.north)[above]{$\theta_m=20\degree$};
\node(PHI030) at (C030.east)[right]{$\varphi= 30\degree$};
\node(PHI050) at (C050.east)[right]{$\varphi= 50\degree$};
\node(PHI070) at (C070.east)[right]{$\varphi= 70\degree$};
\node(PHI090) at (C090.east)[right]{$\varphi= 90\degree$};
\node(PHI110) at (C110.east)[right]{$\varphi=110\degree$};
\node(PHI130) at (C130.east)[right]{$\varphi=130\degree$};
\setcounter{myplotslabel}{1}
\myoplabelar{A030}{1.5715in}{1.0094in}
\myoplabelar{B030}{1.5715in}{1.0094in}
\myoplabelar{C030}{1.5715in}{1.0094in}
\myoplabelar{A050}{1.5715in}{1.0094in}
\myoplabelar{B050}{1.5715in}{1.0094in}
\myoplabelar{C050}{1.5715in}{1.0094in}
\myoplabelar{A070}{1.5715in}{1.0094in}
\myoplabelar{B070}{1.5715in}{1.0094in}
\myoplabelar{C070}{1.5715in}{1.0094in}
\myoplabelar{A090}{1.5715in}{1.0094in}
\myoplabelar{B090}{1.5715in}{1.0094in}
\myoplabelar{C090}{1.5715in}{1.0094in}
\myoplabelar{A110}{1.5715in}{1.0094in}
\myoplabelar{B110}{1.5715in}{1.0094in}
\myoplabelar{C110}{1.5715in}{1.0094in}
\myoplabelar{A130}{1.5715in}{1.0094in}
\myoplabelar{B130}{1.5715in}{1.0094in}
\myoplabelar{C130}{1.5715in}{1.0094in}
\end{tikzpicture}
\caption{Total lift obtained from the 
DNS (\lineSymbolRGB[solid]{0}{0}{0}{0.012in}{none}{0in}) together with the 
estimations with M1 (\lineSymbolRGB[solid]{1}{0}{0}{0.012in}{none}{0in}) 
and M2 (\lineSymbolRGB[solid]{0}{0}{1}{0.012in}{none}{0in}).
\label{fig:totcl}}
\end{center}
\end{figure}
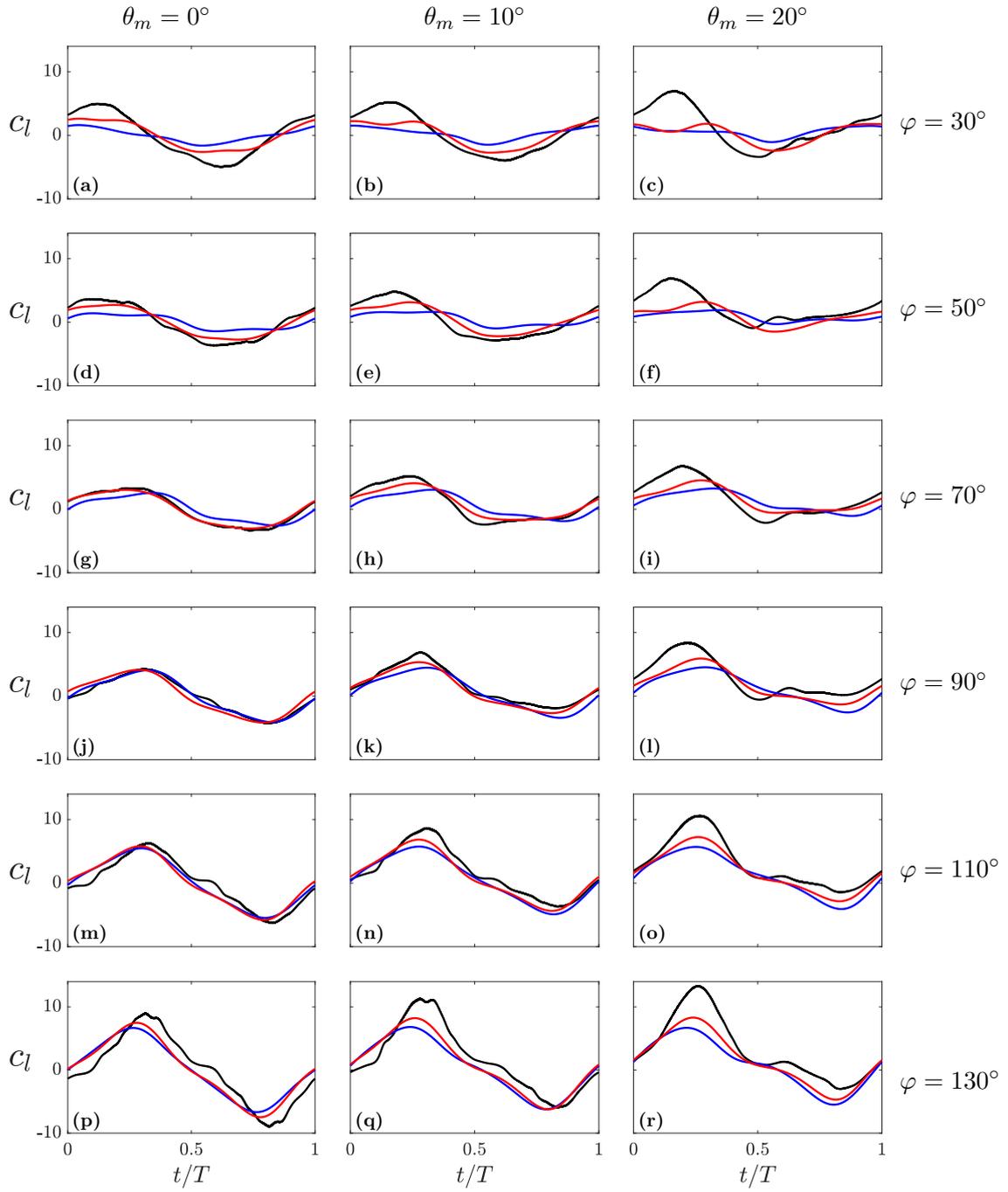

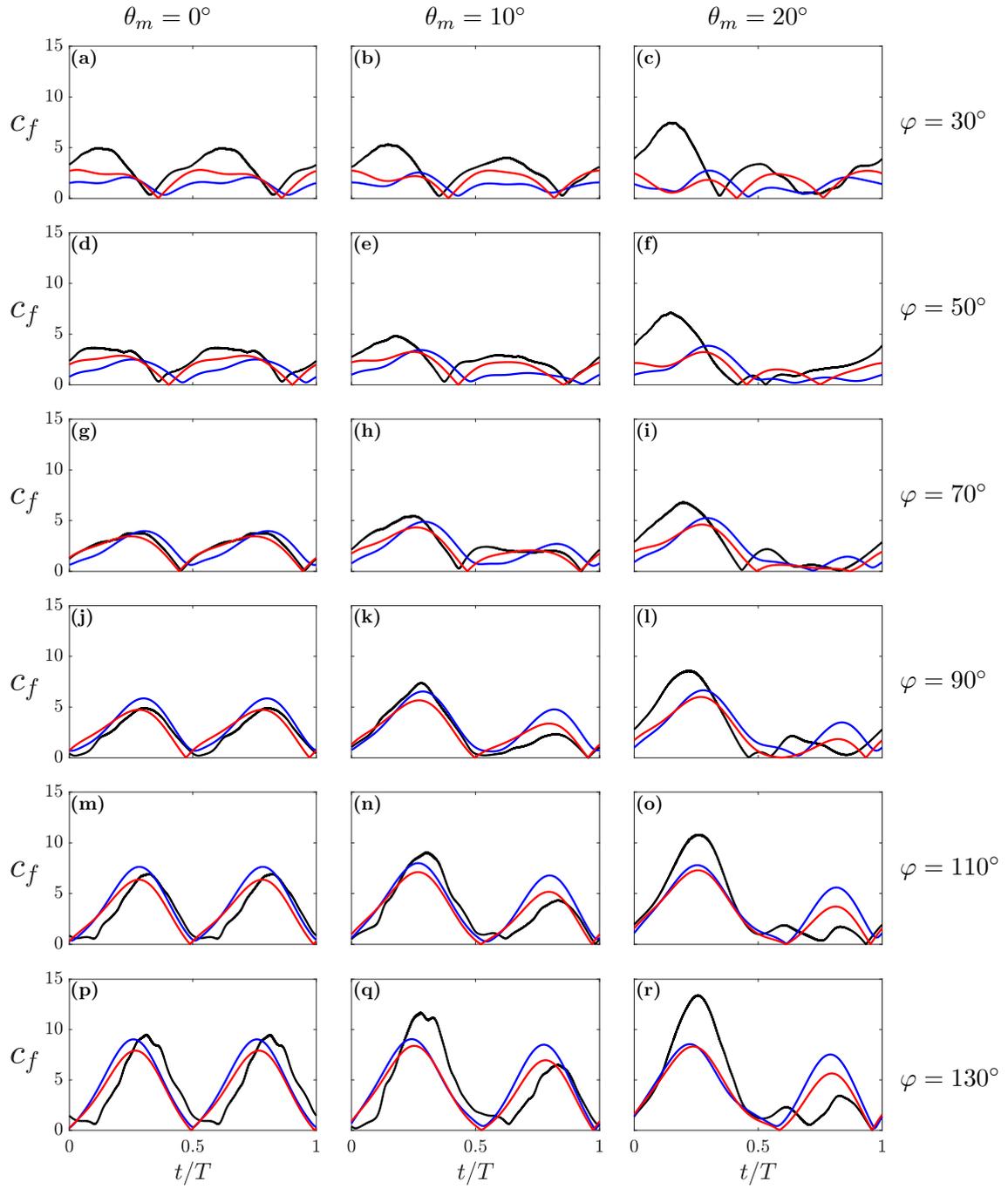
\begin{figure}
\begin{center}
\begin{tikzpicture}
\coordinate(O) at (0.,0.);
\node(A030) at (O)                {\ig[scale=0.5]{fig15a.eps}};
\node(B030) at (A030.east)[right] {\ig[scale=0.5]{fig15b.eps}};
\node(C030) at (B030.east)[right] {\ig[scale=0.5]{fig15c.eps}};
\node(A050) at (A030.south)[below]{\ig[scale=0.5]{fig15d.eps}};
\node(B050) at (B030.south)[below]{\ig[scale=0.5]{fig15e.eps}};
\node(C050) at (C030.south)[below]{\ig[scale=0.5]{fig15f.eps}};
\node(A070) at (A050.south)[below]{\ig[scale=0.5]{fig15g.eps}};
\node(B070) at (B050.south)[below]{\ig[scale=0.5]{fig15h.eps}};
\node(C070) at (C050.south)[below]{\ig[scale=0.5]{fig15i.eps}};
\node(A090) at (A070.south)[below]{\ig[scale=0.5]{fig15j.eps}};
\node(B090) at (B070.south)[below]{\ig[scale=0.5]{fig15k.eps}};
\node(C090) at (C070.south)[below]{\ig[scale=0.5]{fig15l.eps}};
\node(A110) at (A090.south)[below]{\ig[scale=0.5]{fig15m.eps}};
\node(B110) at (B090.south)[below]{\ig[scale=0.5]{fig15n.eps}};
\node(C110) at (C090.south)[below]{\ig[scale=0.5]{fig15o.eps}};
\node(A130) at (A110.south)[below]{\ig[scale=0.5]{fig15p.eps}};
\node(B130) at (B110.south)[below]{\ig[scale=0.5]{fig15q.eps}};
\node(C130) at (C110.south)[below]{\ig[scale=0.5]{fig15r.eps}};
% labels
\node(TH00) at (A030.north)[above]{$\theta_m= 0\degree$};
\node(TH10) at (B030.north)[above]{$\theta_m=10\degree$};
\node(TH20) at (C030.north)[above]{$\theta_m=20\degree$};
\node(PHI030) at (C030.east)[right]{$\varphi= 30\degree$};
\node(PHI050) at (C050.east)[right]{$\varphi= 50\degree$};
\node(PHI070) at (C070.east)[right]{$\varphi= 70\degree$};
\node(PHI090) at (C090.east)[right]{$\varphi= 90\degree$};
\node(PHI110) at (C110.east)[right]{$\varphi=110\degree$};
\node(PHI130) at (C130.east)[right]{$\varphi=130\degree$};
\setcounter{myplotslabel}{1}
\myoplabelbr{A030}{1.5715in}{0.12in}
\myoplabelbr{B030}{1.5715in}{0.12in}
\myoplabelbr{C030}{1.5715in}{0.12in}
\myoplabelbr{A050}{1.5715in}{0.12in}
\myoplabelbr{B050}{1.5715in}{0.12in}
\myoplabelbr{C050}{1.5715in}{0.12in}
\myoplabelbr{A070}{1.5715in}{0.12in}
\myoplabelbr{B070}{1.5715in}{0.12in}
\myoplabelbr{C070}{1.5715in}{0.12in}
\myoplabelbr{A090}{1.5715in}{0.12in}
\myoplabelbr{B090}{1.5715in}{0.12in}
\myoplabelbr{C090}{1.5715in}{0.12in}
\myoplabelbr{A110}{1.5715in}{0.12in}
\myoplabelbr{B110}{1.5715in}{0.12in}
\myoplabelbr{C110}{1.5715in}{0.12in}
\myoplabelbr{A130}{1.5715in}{0.12in}
\myoplabelbr{B130}{1.5715in}{0.12in}
\myoplabelbr{C130}{1.5715in}{0.12in}
\end{tikzpicture}
\caption{Total aerodynamic force modulus obtained from the 
DNS (\lineSymbolRGB[solid]{0}{0}{0}{0.012in}{none}{0in}) together with the 
estimations with M1 (\lineSymbolRGB[solid]{1}{0}{0}{0.012in}{none}{0in}) 
and 
M2 (\lineSymbolRGB[solid]{0}{0}{1}{0.012in}{none}{0in}).
\label{fig:totcc}}
\end{center}
\end{figure}

%-------------------------------------------------------------------------------
Table \ref{tab:total} shows a quantitative comparison of the estimations of M1
and M2 in terms of the differences between mean and instantaneous values of the
lift and thrust coefficients of each model and the DNS.
The data in table \ref{tab:total} are complemented by figures \ref{fig:totct},
\ref{fig:totcl} and \ref{fig:totcc}, which show the instantaneous force
coefficients obtained with both models and with the DNS.  
%_______________________________________________________________________________

%-------------------------------------------------------------------------------
Starting with the thrust, predicted mean and instantaneous values are 
consistently improved with M1 with respect to M2.
Furthermore, $\overline{c}_{t,M1}$ shows good agreement with the DNS results, 
with differences between M1 and DNS lower than $20\%$ of the rms of the total
thrust.
The estimate of mean thrust becomes less accurate for cases with $\theta_m=20
\degree$, with differences of the order of $50\%$ of the rms of the total force. 
Interestingly, cases with $\pphi = 70\degree$ show the smallest differences in
thrust for each $\theta_m$. 
This can also be appreciated in the instantaneous values of $c_t$ shown in
figure \ref{fig:totct} (compare the results of the third row to the other
panels).  
As with the mean values, the estimations of the instantaneous thrust
coefficients of M1 are less accurate for $\theta_m = 20\degree$.
Also, when $\varphi \ge 110\degree$, the instantaneous values of $c_t$ from M1
tend to peak earlier than the corresponding DNS values (see figures
\ref{fig:totct}m, n, p and q).   
%_______________________________________________________________________________

%-------------------------------------------------------------------------------
%-------------------------------------------------------------------------------

\def\mf{\ref{fig:totcl}}
%-------------------------------------------------------------------------------
Regarding the mean lift, the predictions of M1 are, in general, better than
those of M2, except for cases B030, B050 and C030.
However, for these cases the $L_2$ norm of the differences between M1 and the
DNS are smaller than those corresponding to M2 (see also figures \mf{}b, cand
e).
The predictions of M1 in terms of the rms of the lift coefficient improve with 
respect to M2 for all cases, except for A090 and A110.
Cases with  phase shift $\pphi<90\degree$ present differences in mean lift 
lower than $5\%$ of the rms of total lift and cases with higher phase shift 
$\pphi\ge 90\degree$, lower than $20\%$.
Instantaneous values are very well captured for cases with $\thm=0\degree,10
\degree$ and $50\le\pphi\le110\degree$ (figure \mf{}).
Again, a shift between both models and the DNS appears for the cases with higher
phase shift ($\pphi=130\degree$), and errors tend to increase when the mean
pitch angle is increased to $\thm=20\degree$.
For the latter, the peak in the lift during the downstroke is consistently
underestimated by both models.
Finally, for the symmetric cases ($\theta_m=0\degree$), the estimated mean lift
is zero in both models (as expected). 
The differences that appear in cases A030 and A130 between the models and the
DNS arise from the averaging process of the DNS data, due to the aperiodic
nature of these cases.
%-------------------------------------------------------------------------------

%-------------------------------------------------------------------------------
%_______________________________________________________________________________

%-------------------------------------------------------------------------------
The different behavior of the models in the prediction of lift and thrust
coefficients are in part due to the ability of the models to predict properly
the orientation of the aerodynamic force.
In order to compare the two models and the DNS without taking into account this
effect, figure \ref{fig:totcc} shows the total aerodynamic force coefficient,
defined as 
\begin{equation} 
c_f = \sqrt{c_l^2 + c_t^2}. 
\end{equation}
It can be observed in figure \ref{fig:totcc} that the agreement between both
models and the DNS for $\theta_m = 0\degree, 10\degree$ and $50\degree
\lesssim \varphi \lesssim 110\degree$ is reasonably good. 
For $\varphi = 30\degree$ and $\theta_m=0^\circ,10^\circ$, the amplitude of
$c_f$ from the DNS is larger than both models. 
For $\varphi = 130\degree$ and $\theta_m=0^\circ,10^\circ$, the magnitudes of
the peaks of $c_f$ are reasonably captured, but the models are shifted in time
with respect the DNS data. 
For $\theta_m= 20\degree$, independently of the value of $\varphi$, the peak of
$c_f$ during the downstroke is underestimated in both models, which suggests
that the model used for the circulation (namely equation \ref{eq:wangmodel})
might not be properly capturing the development and evolution of the LEV for
these high angle of attack cases. 
Note also that the differences between M1 and M2 in $c_f$ and $c_l$ (figures
\ref{fig:totcl} and \ref{fig:totcc}) are less apparent than in $c_t$ in figure
\ref{fig:totct}.
This suggests that the chord normal orientation of the contribution of the
vorticity within the flow is more important for the prediction of the thrust. 

Finally, note that the model proposed here has some limitations.
As discussed in section \ref{sec:ref}, $\vec{F}^v$ is approximately normal to
the chord of the airfoil when the LEV is strong enough, and even in that case
there is a small component of $\vec{F}^v$ tangential to the airfoil chord.
This tangential component is unimportant for the cases discussed here, but it
can be significant in some cases, like the pure heaving case.
Indeed, for a pure heaving case at zero pitch angle, M1 predicts $c_t=0$ (since
both vorticity within the flow and body motion components result in forces
perpendicular to the chord).
This means that further work is needed to widen the range of applicability of
M1, in order to include pure heaving cases.
%_______________________________________________________________________________

%

%=============================================conclusions=======================
%=============================================conclusions=======================
%=============================================conclusions=======================
\section{Conclusions}
\label{sec:concl}

In this work we have generated and analyzed a database of DNS of heaving and
pitching airfoils, with large amplitude motions, moderate Reynolds numbers and
reduced frequencies of order $O(1)$.
In order to analyze the effect of the mean pitch angle and the phase shift, we
have varied these parameters in the ranges $\thm{} \in [0\degree,20\degree]$
and $\pphi \in [30\degree,130\degree]$.
In terms of the lift and thrust obtained by these configurations, we observe 
that as \thm{} increases the mean lift increases and the mean thrust decreases
with little variations in the amplitude of the instantaneous fluctuations.
This is consistent with a change in the orientation of the force when \thm{} 
varies, analogous to a change in the stroke plane.
On the other hand, the effect of the phase shift is less intuitive: 
the net lift and thrust increase with \pphi{} for $\pphi \lesssim 110\degree$,
as well as the amplitude of their fluctuations.
As a consequence, in our database the maximum propulsion efficiency is obtained 
for $\pphi = 90\degree$ and the maximum force is obtained for $\pphi= 
110\degree$.
This is consistent with previous investigations \citep{anderson:1998,
ramamurti:2001}.
%_______________________________________________________________________________

%  FORCE DECOMPOSITION
%-------------------------------------------------------------------------------
Subsequently, we have decomposed the total aerodynamic force following 
\citet{chang1992potential}.
We observe that for the considered kinematics, the contributions from the
vorticity within the flow and body motion are comparable, while the surface
vorticity contribution is significantly smaller.
The contribution of vorticity within the flow is influenced only by the vortices
in the vicinity of the airfoil and is roughly perpendicular to the chord.
Note that since vorticity within the flow is the main contributor to the net 
aerodynamic forces, the effects of \pphi{} and \thm{} are the same as discussed
in the previous paragraph for the total forces.
Conversely, for a given value of \thm{}, the net contribution from body motion 
to the thrust and lift increases monotonically with \pphi{}.
%_______________________________________________________________________________

%  MODEL
%-------------------------------------------------------------------------------
Finally, based on the observations made from the force decomposition analysis,
we have proposed and tested a reduced order model for the aerodynamic forces.
We compute the body motion contribution directly from 
\citet{chang1992potential}, since it only depends on geometric characteristics
of the airfoil (the auxiliary potentials) and on the kinematics.
We model the vorticity within the flow contribution as a force perpendicular to
the chord and whose modulus is proportional to the circulation of the airfoil.
This circulation is calculated with the model of \citet{pesavento2004falling}.
The only parameters that need fitting are the constants $G_T$ and $G_R$
appearing in the model for the circulation.
Our results show that these two constants can be fixed for the whole database
keeping a reasonable error in the force estimation, which suggests that the
model is able to take into account the kinematic parameters that are varied in
this study (\thm{} and \pphi{}). 
Overall, the model is able to predict mean thrust and lift with errors smaller
than $20\%$ of the rms of the corresponding forces for $\thm=0\degree$ and 
$10\degree$.
From the point of view of the instantaneous forces the agreement between the 
model and the DNS data is very satisfactory for $\thm = 0\degree$ and
$10\degree$, and $50 \degree \lesssim \pphi \lesssim 110\degree$.
For $\thm = 20\degree$, the instantaneous forces suggest that the error in the
estimation of the mean forces is due to the underestimation of the intensity of
the LEV during the downstroke.
The predictions of the proposed model are compared to those of a similar model
from the literature, showing a noticeable improvement on the prediction of the
mean thrust, and a smaller improvement on the prediction of mean lift and the
instantaneous force coefficients.\\
%_______________________________________________________________________________

% ACKNOWLEDGMENTS
%-------------------------------------------------------------------------------
The authors acknowledge the support received by the Spanish Ministry of 
Economy and Competitiveness through grant TRA2013-41103-P.
This grant includes FEDER funding.
%_______________________________________________________________________________

\bibliography{./bibdata}

\appendix

%_______________________________________________________________________________
% REBUTTAL BEG
\section{Grid refinement study}
\label{sec:A1}

%-------------------------------------------------------------------------------
%
\begin{figure}
\begin{center}
\begin{tikzpicture}
\coordinate(O) at (0.,0.);
\node(A) at (O)             {\ig[scale=0.5]{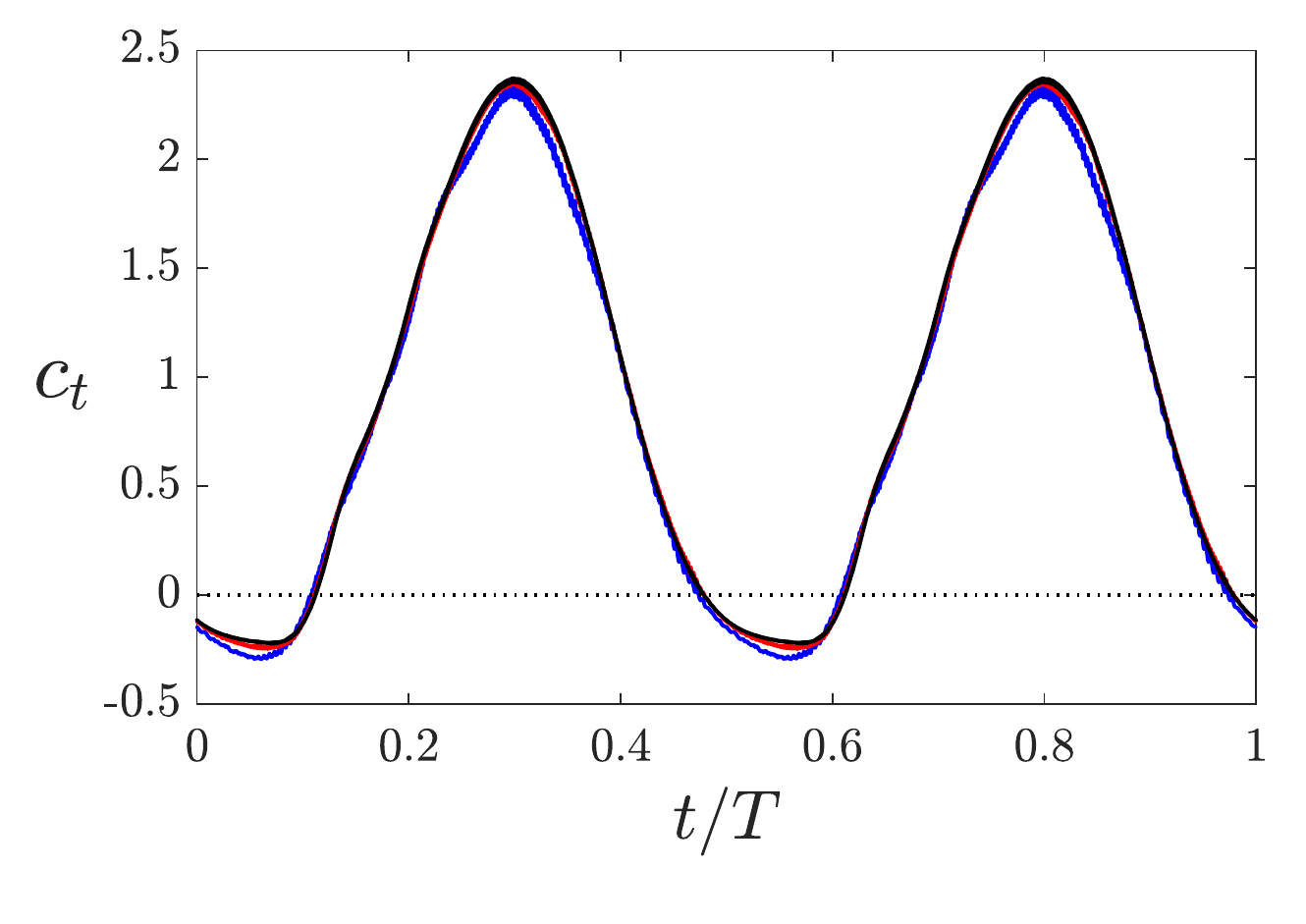}};
\node(B) at (A.east)[right] {\ig[scale=0.5]{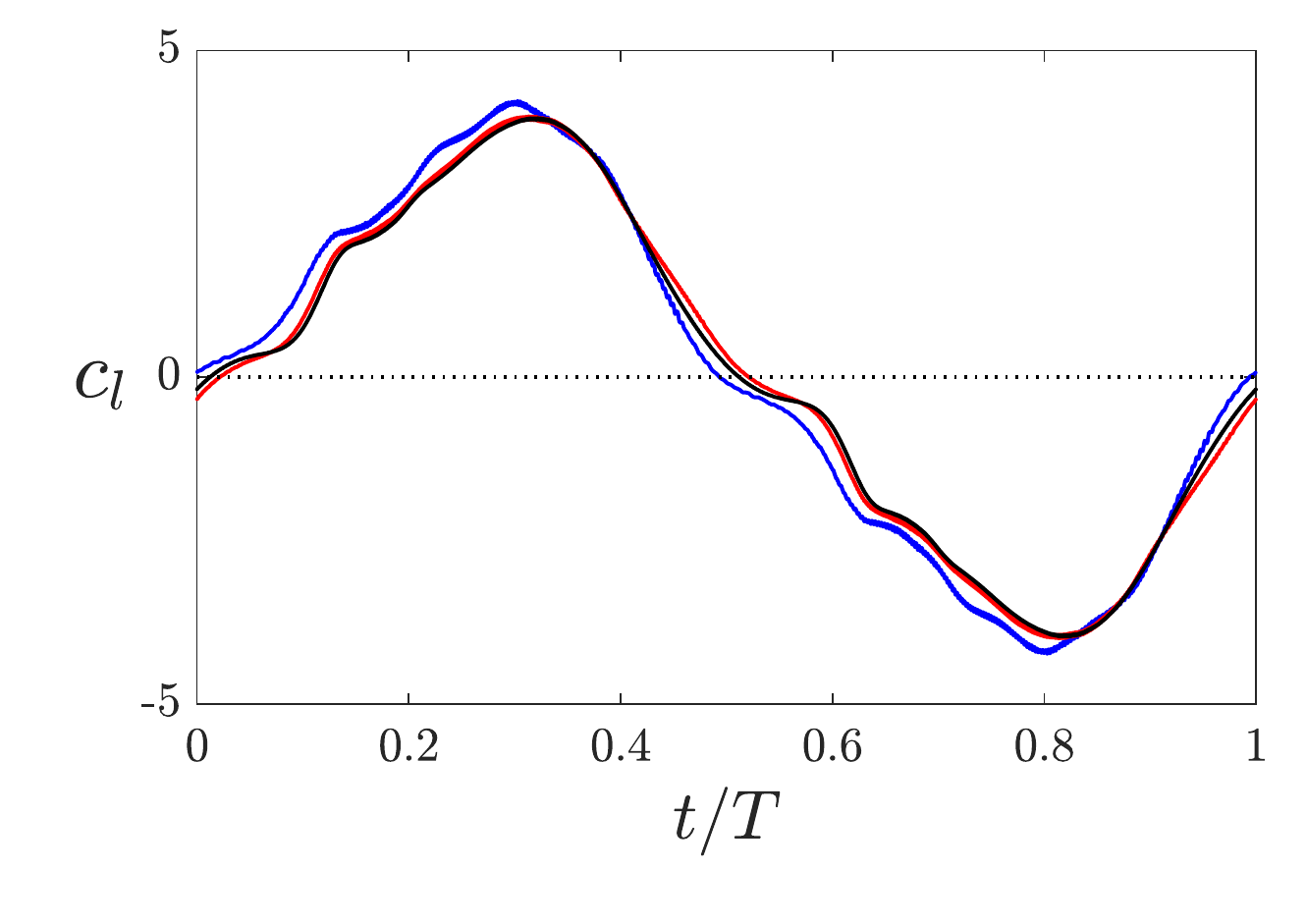}};
\node(C) at (A.south)[below]{\ig[scale=0.5]{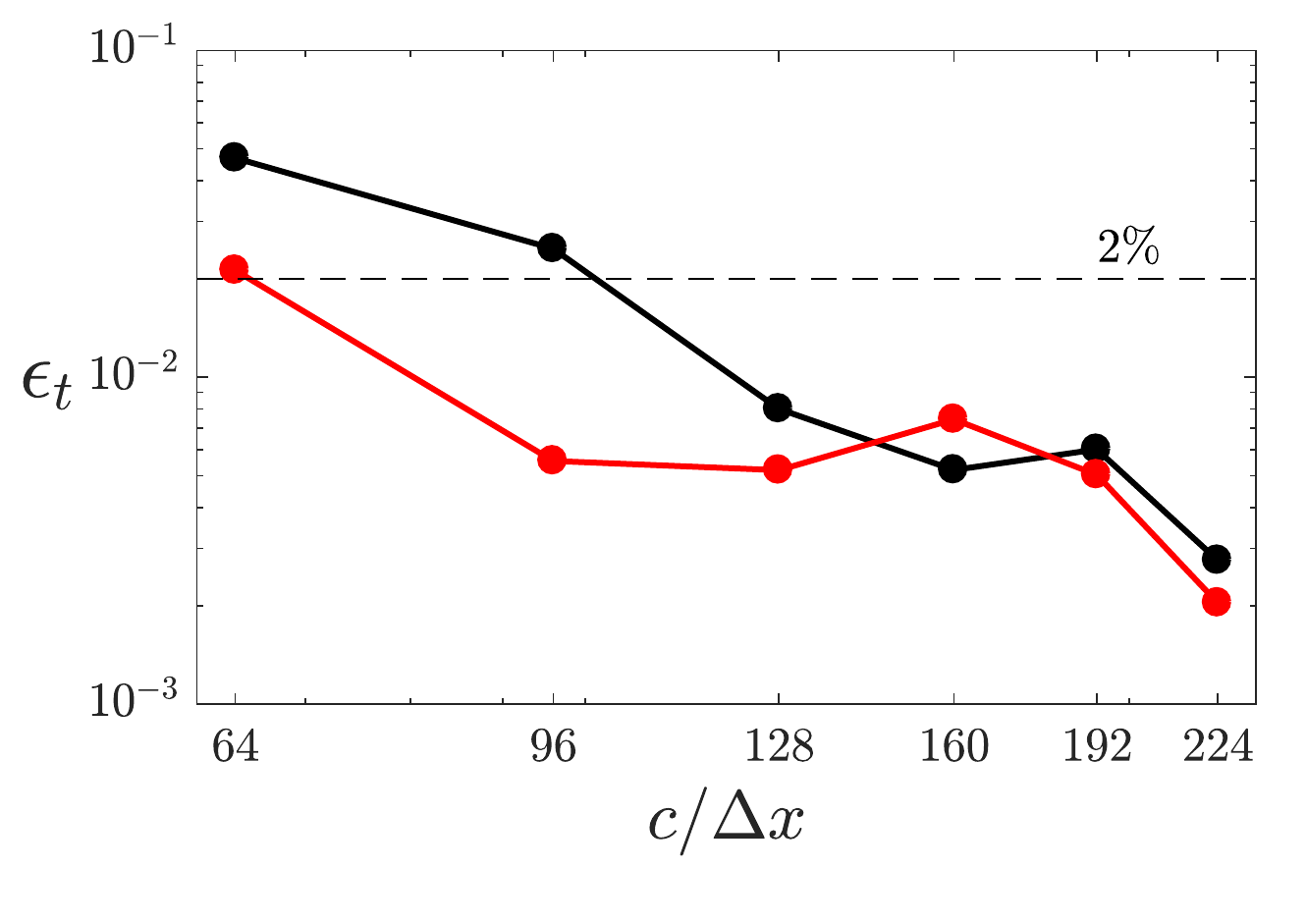}};
\node(D) at (C.east)[right] {\ig[scale=0.5]{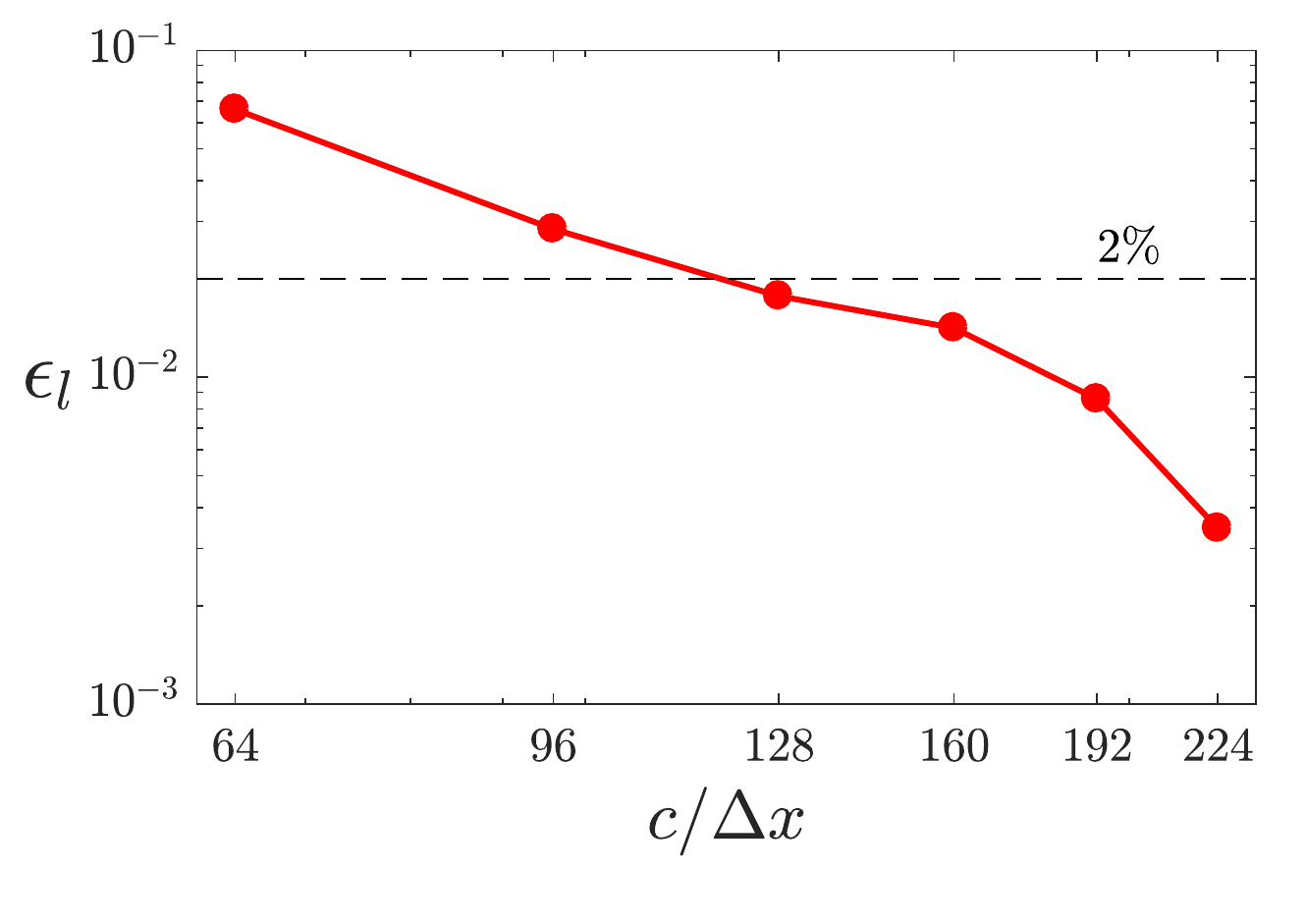}};
\setcounter{myplotslabel}{1}
\myoplabelar{A}{2.0572in}{1.4332in}
\myoplabelar{B}{2.2572in}{1.4332in}
\myoplabelar{C}{2.2572in}{1.4332in}
\myoplabelar{D}{2.2572in}{1.4332in}
\end{tikzpicture}
\caption{Evolution of a) thrust and b) lift for cases with 
$c/\Delta x = 64$ (\lineSymbolRGB[solid]{0}{0}{1}{0.012in}{none}{0.048in}),
$c/\Delta x =128$ (\lineSymbolRGB[solid]{1}{0}{0}{0.012in}{none}{0.048in}) and
$c/\Delta x =256$ (\lineSymbolRGB[solid]{0}{0}{0}{0.012in}{none}{0.048in}).
Errors obtained in 
mean (\lineSymbolRGB[solid]{0}{0}{0}{0.012in}{o*}{0.048in}) and 
rms  (\lineSymbolRGB[solid]{1}{0}{0}{0.012in}{o*}{0.048in}) values of c) thrust 
and d) lift of all the simulations.
\label{fig:A1/err}}
\end{center}
\end{figure}
In this section we present the grid refinement study carried out to select the 
resolution used in the simulations.
In order to save computational time, a smaller computational domain has been
employed in the grid refinement study, namely, $12c \times 8c$ in the streamwise
and vertical directions, respectively.
The parameters of the case under consideration correspond to case A090 reported
in table \ref{tab:clct_all}.
This is a case in which mean thrust is produced while the mean lift is zero.
We have performed seven simulations varying the resolution from $c/\Delta x=64$
to $c/\Delta x = 256$ and the time step $\Delta t$ is set accordingly to keep a 
CFL number lower than $0.2$.
We study the convergence of the aerodynamic forces when increasing the
resolution, and we use as a reference data the results of the case with the
highest resolution ($c/\Delta x = 256$).
The time evolution of the thrust and lift coefficients of three of the
simulations are reported in figure \ref{fig:A1/err}a and b.
While some deviations are observed for the resolution $c/\Delta x = 64$ with 
respect to the reference case, the results of the simulation with resolution 
$c/\Delta x = 128$ are very close to those of the reference case.
In order to quantify the differences we define the errors in the mean and rms
of the forces
\begin{subequations}\label{eq:A/err}
\begin{align}
\epsilon_t^\text{mean}(r)&= \frac{|\ctm_r-\ctm_{256}|}{\ctr_{256}}, \\
\epsilon_t^\text{rms}(r) &= \frac{|\ctr_r-\ctr_{256}|}{\ctr_{256}}, \\
\epsilon_l^\text{mean}(r)&= \frac{|\clm_r|}{\clr_{256}}, \\
\epsilon_l^\text{rms}(r) &= \frac{|\clr_r-\clr_{256}|}{\clr_{256}},
\end{align}
\end{subequations}
where $r$ is the resolution of each case.
Since the case under consideration produces no lift, the definition of
$\epsilon_l^\text{mean}(r)$ does not need to have a reference mean lift.
Figures \ref{fig:A1/err}c and d show the errors as a function of the resolution.
Note that $\epsilon_l^\text{mean}(r)$ is not shown in figure \ref{fig:A1/err}d
since it is smaller than $10^{-4}$ for all the cases. 
In general the errors decrease with increasing resolution.
Taking into account a compromise between the computational cost and the accuracy 
of the results, we have decided to use a resolution of $c/\Delta x = 128$ in our 
calculations.
With this resolution the errors for both mean and rms of the forces are smaller 
than $2\%$ as shown in the figure.
%_______________________________________________________________________________

\section{Calculation of auxiliary potential functions}
\label{sec:rotation}

%-------------------------------------------------------------------------------
The force decomposition algorithm used in  this  work was first introduced by 
 \cite{chang1992potential} and later used by 
\cite{martin2015vortex}.
The algorithm requires two auxiliary potential functions \phix{} and \phiz{} 
(see equation \ref{eq:force_decomp}), which are defined as
\begin{subequations}\label{eq:phix}
\begin{align}
\nabla^2\phix           & =0 \\
\nabla\phix\cdot\vec{n} & =-\vec{n}\cdot\ex\,U_\infty & \text{At the body surface}
                                              \label{eq:phix_bcbody}\\
\phix                   & \rightarrow 0     & \text{ At infinity,}
\end{align}
\end{subequations}

\begin{subequations}\label{eq:phiz}
\begin{align}
\nabla^2\phiz           & =0 \\
\nabla\phiz\cdot\vec{n} & =-\vec{n}\cdot\ez\,U_\infty & \text{At the body surface}
                                              \label{eq:phiz_bcbody}\\
\phiz                   & \rightarrow 0     & \text{ At infinity,}
\end{align}
\end{subequations}
where $\vec{n}$ is a unitary vector normal to the surface of the airfoil 
pointing towards the fluid.
%_______________________________________________________________________________

%-------------------------------------------------------------------------------
The auxiliary potential functions \phix{} and \phiz{} are needed at every time 
step in which the force decomposition is to be applied.
These potentials depend only on the airfoil geometry and on the direction of the
free stream velocity, not on the heaving and pitching velocities and 
accelerations.
Note that the dependency with the orientation is linear, so that it is possible 
to show that for an arbitrary direction $\vec{\alpha}= \alpha_1\ex+\alpha_2\ez$,
the corresponding potential $\phi_\alpha$ satisfies
\begin{subequations}\label{eq:alpha}
\begin{align}
\nabla^2\phi_\alpha           & =0 \\
\nabla\phi_\alpha\cdot\vec{n} & =-\vec{n}\cdot\vec{\alpha}\,U_\infty & \text{At the body surface}
                                              \label{eq:alpha_bcbody}\\
\phi_\alpha                   & \rightarrow 0     & \text{ At infinity}. 
\end{align}
\end{subequations}
Hence this potential can be computed as 
$\phi_\alpha=\alpha_1\phi_x+\alpha_2 \phi_z$.

%-------------------------------------------------------------------------------
The linearity with the orientation eliminates the problem of having to compute
the auxiliary potential functions at different time instants.
Instead, the auxiliary potential functions are computed for a reference position
of the airfoil, and then rotated (using \ref{eq:alpha}) and translated to the
position of the airfoil at each time instant. 
%_______________________________________________________________________________

\begin{figure}
\begin{center}
\begin{tikzpicture}
\coordinate(O) at (0.,0.);
\node(A) at (O)            {\ig[scale=0.5]{fig17a.eps}};
\node(B) at (A.east)[right]{\ig[scale=0.5]{fig17b.eps}};
\setcounter{myplotslabel}{1}
\mylabel{A}{0.3in}{0.3in}
\mylabel{B}{0.3in}{0.3in}
\end{tikzpicture}
\caption{Auxiliary potentials a) \phix{} and b) \phiz{} for a NACA 0012 
airfoil at $\theta = 0\degree$.
\label{fig:phiatfoil}}
\end{center}
\end{figure}
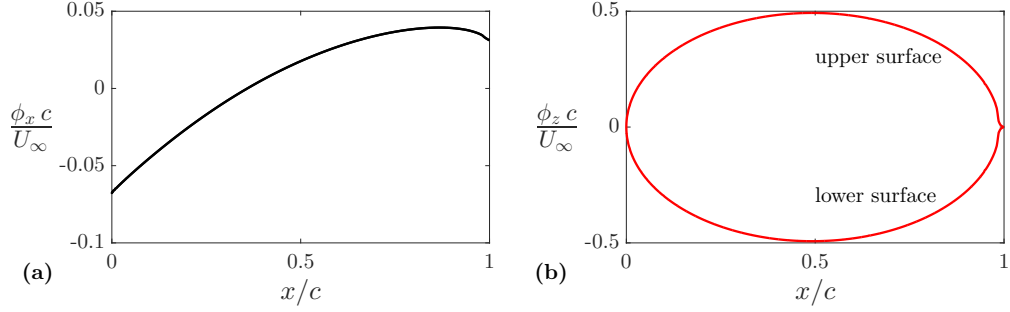

%-------------------------------------------------------------------------------
Finally, it should be noted that the auxiliary potential functions \phix{} and
\phiz{} only have analytical solutions for simple geometries, like ellipses
\citep{martin2015vortex}. 
For other geometries, like the NACA 0012 considered in the present work, they 
need to be computed numerically. 
This is done using the sharp interface method of \cite{mittal2008versatile}, 
which is based on an immersed boundary formulation where normal derivatives on 
the solid boundary are imposed by using image and ghost points. 
The implementation of the solver has been validated computing the potential
function for ellipses of different aspect ratios, and comparing the numerical
results to the analytical solutions. 
For the computations of the potential functions at the reference orientation of 
the NACA 0012 airfoils (shown in figure \ref{fig:phiatfoil}), we have employed 
a computational domain of $30c$ x $30c$. This domain has been discretized using 
$6072$ x $6072$ grid points in the chordwise and vertical directions, 
respectively. 
The translated and rotated potential functions are then interpolated with a 
linear interpolator to the collocation points where the integrals of
\eqref{eq:force_decomp} are computed.
%_______________________________________________________________________________

%-------------------------------------------------------------------------------
In order to compute the body motion contribution $\vec{F}^m$ to the total 
aerodynamic force, we present the value of the potentials \phix{} and \phiz{} at
the surface of a NACA 0012 airfoil in the reference position (figures 
\ref{fig:phiatfoil}a and b).
%_______________________________________________________________________________

\end{document}